\documentclass[journal]{IEEEtran}
\usepackage{widetext}
\ifCLASSINFOpdf
  \usepackage[pdftex]{graphicx}
  \graphicspath{{../pdf/}{../jpeg/}}
  \DeclareGraphicsExtensions{.pdf,.jpeg,.png}
\else
  \usepackage[dvips]{graphicx}
  \graphicspath{{../eps/}}
  \DeclareGraphicsExtensions{.eps}
\fi
\usepackage{amsmath}
\usepackage{amsfonts}
\usepackage{bm}

\usepackage{subcaption}

\begin{document}
%
\title{A Volumetric Approach to Point Cloud Compression}
%
%
%

\author{Maja~Krivoku\'ca, 
        Maxim~Koroteev, 
      and~Philip~A.~Chou 
}

%
%

\markboth{8i Technical Report, September 2018}
{
A Volumetric Approach to Point Cloud Representation and Compression}
%



\maketitle

\begin{abstract}
Compression of point clouds has so far been confined to coding the positions of a discrete set of points in space and the attributes of those discrete points.  We introduce an alternative approach based on volumetric functions, which are functions defined not just on a finite set of points, but throughout space.  As in regression analysis, volumetric functions are continuous functions that are able to interpolate values on a finite set of points as linear combinations of continuous basis functions.  Using a B-spline wavelet basis, we are able to code volumetric functions representing both geometry and attributes.  Geometry is represented implicitly as the level set of a volumetric function (the signed distance function or similar).  Attributes are represented by a volumetric function whose coefficients can be regarded as a critically sampled orthonormal transform that generalizes the recent successful region-adaptive hierarchical (or Haar) transform to higher orders.  Experimental results show that both geometry and attribute compression using volumetric functions improve over those used in the emerging MPEG Point Cloud Compression standard.
\end{abstract}

\begin{IEEEkeywords}
B\'ezier volumes, B-splines, wavelets, point cloud compression, geometry coding, shape coding, color coding, attribute coding, multiresolution representations, signed distance function, graph signal processing, counting measure.
\end{IEEEkeywords}

%
\IEEEpeerreviewmaketitle

\section{Introduction}
%
%
%
%

\begin{figure}[!t]
\centering
\includegraphics[width=0.65\linewidth]{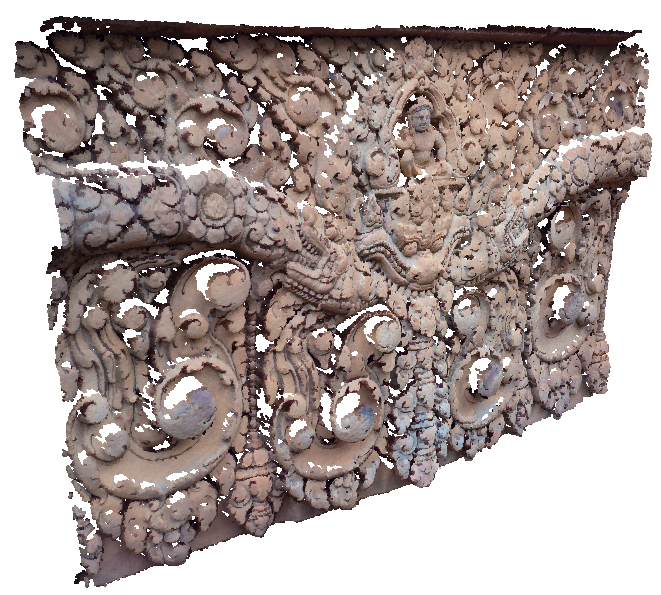}
\includegraphics[width=0.3\linewidth]{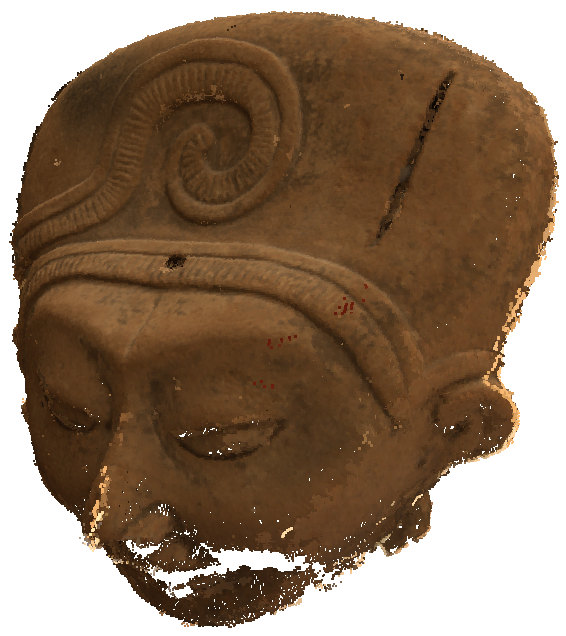}
\includegraphics[width=0.85\linewidth]{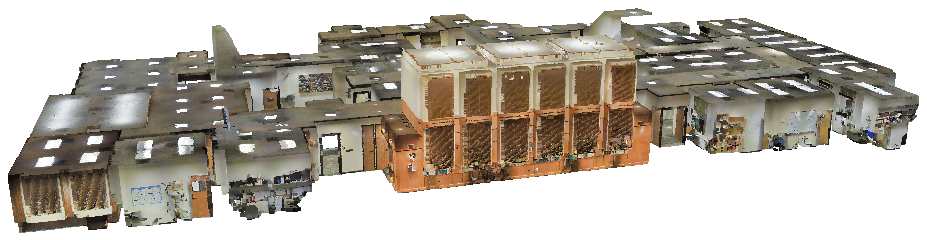}
\includegraphics[width=0.65\linewidth]{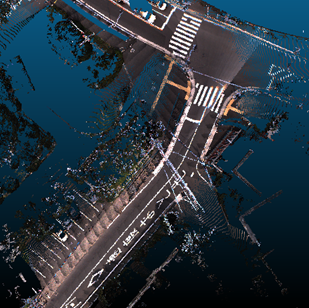}
\includegraphics[width=0.23\linewidth]{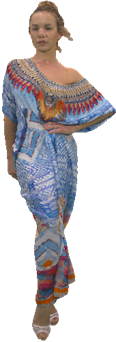}
\caption{Examples of point clouds: facades, cultural artifacts, buildings, cities, and people.}
\label{fig:examplePointClouds}
\end{figure}

\IEEEPARstart{A}{\em point} {\em cloud} is a set of {\em points} (also known as {\em locations} or {\em positions}) in Euclidean space, in which a vector of attributes (such as a color triple) may be associated with each point.
Point clouds may be {\em static} or {\em dynamic}.
A dynamic point cloud is a sequence of static point clouds, each in its own {\em frame}.
Figure~\ref{fig:examplePointClouds} shows several examples of point clouds.  Point clouds have applications in robotics, tele-operation, virtual and augmented reality, cultural heritage preservation, geographic information systems, and so forth.

Point clouds offer a natural representation of {\em volumetric media}, popularly known as {\em holograms}.
Broadly interpreted, a hologram is an object or scene whose representation permits rendering arbitrary points of view, such that the object or scene appears to occupy space, due to stereo or motion parallax.  Point clouds offer a natural representation of holograms, because each point, with one or more color attributes, can naturally represent the color of rays that pass through that point.

Volumetric media, or holograms, have emerged as the first significant new modality for immersive communication since the introduction in the late nineteenth century of audio for audio recordings and video for motion pictures.
Like audio and video, volumetric media will be used in three major communication scenarios: on-demand consumption of pre-recorded content,  broadcast of live or pre-recorded content, and interactive communication such as telephony or conferencing.
For these scenarios, storage and transmission are essential; hence data compression is essential.

Interest in point cloud compression has been demonstrated in recent publications, as well as in the establishment of a standardization activity within MPEG (see next section).

The two fundamental problems in point cloud compression are {\em geometry} compression and {\em attribute} compression.  Geometry compression, which is sometimes called {\em shape} compression, is the problem of compressing the point locations; attribute compression is the problem of compressing the attribute values, given the point locations.

In this paper, we introduce a volumetric approach to point cloud compression, which can be used for both geometry and attribute compression.  In our volumetric approach, both geometry and attributes are represented by volumetric functions.  We use the term {\em volumetric function} to refer to a scalar or vector valued function defined on a volume of space (in contrast, say, to a function defined on an image plane or on a finite set of points).  To create a compressed representation of the geometry or attributes, the appropriate volumetric functions are transformed, quantized, and entropy coded.  Once decoded, the decoded volumetric functions are used to reconstruct the geometry and attributes.

To represent geometry, we use a scalar volumetric function and reconstruct the geometry as the level set of the decoded volumetric function.  As the volumetric function for geometry, we typically use a signed distance function or an occupancy probability.

To represent the attributes, we use a vector-valued volumetric function having the same dimension as the attributes, and we reconstruct the attributes as the values of the decoded volumetric function at the points of the decoded geometry.  We determine the volumetric function by solving a linear regression.  The parameters of the function are B-spline wavelet coefficients, equivalent to the parameters of linked B\'ezier volumes.  The functions exist in a Hilbert space in which norm and orthogonality are induced by an inner product defined by a novel counting measure supported by the decoded point locations.

This paper is organized as follows.  Section~\ref{sec:prior} outlines prior work.  Section~\ref{sec:preliminaries} establishes the framework.  Section~\ref{sec:attributes} focuses on attribute coding, while Section~\ref{sec:geometry} focuses on geometry coding.  Section~\ref{sec:results} provides experimental results, and Section~\ref{sec:conclusion} concludes.


\section{Prior Work}
\label{sec:prior}



Much of the prior work on point cloud compression has been rolled up into the emerging MPEG Point Cloud Compression (PCC) standard \cite{Schwarz:18}.  PCC can therefore be used as a template for describing prior work in point cloud compression.  PCC is divided into two profiles, one based on video coding and one based on native 3D coding.  The approach taken in the video-based profile is an adaptation of the ideas of geometry images \cite{GuGH02} and geometry videos \cite{BricenoSMGH03,ColletCSGECHKS15}, which were previously used for mesh compression.  The idea is to flatten patches of colored geometry onto the plane and code them as video, in order to re-use our substantial investment in video coding.  More relevant to the present paper is the native 3D profile of PCC.  In this profile, geometry is coded using octrees and attributes are coded using critically sampled spatial transforms that are adapted to the geometry of each point cloud.  We now give references to prior work in those areas.

Octrees, or more precisely, Sparse Voxel Octrees (SVOs) were developed in the 1980s to represent the geometry of three-dimensional objects \cite{JackinsT80,meagher_octtree}.   In the guise of occupancy grids, they have also had significant use in robotics \cite{Moravec88,Elfes89,PathakBPS07}.  Octrees were first used for point cloud compression in \cite{schnabel_2006}.  They were further developed for progressive point cloud coding, including color attribute compression, in \cite{Huang2008}.  Octrees were extended to coding dynamic point clouds in \cite{Kammerl2012} and implemented in the Point Cloud Library \cite{Rusu_3dis}.  SVOs have been shown to have highly efficient implementations suitable for encoding at video frame rates \cite{Loop_2013}, based on Morton codes \cite{Morton66}.  Entropy coding for octrees has been addressed in \cite{LasserreF:18,QueirozGCF:18,GarciaQ:18}.  In PCC, octrees are pruned at a certain depth, and then represent geometry within each leaf as a single point (i.e., voxel), a list of points, or a geometric model.  The geometric model in PCC are triangulations called a triangle soup.  Polygon soup representations of geometry for point cloud compression were first explored in \cite{PavezCQO16,PavezC:17}.

Point cloud attribute coding using critically sampled spatial transforms adapted to the geometry of each point cloud was first explored in \cite{zhang_icip_2014} using the graph Fourier transform (GFT) \cite{Shuman2013} by considering point cloud attributes as a signal defined on the nodes in a discrete graph corresponding to points of the point cloud.  The GFT was also used for compressing point cloud attributes in \cite{thanou_icip_2015,ThanouCF16,CohenTV16}. Point cloud attribute coding using the Karhunen-Lo\`eve transform (KLT) of a signal defined on points of a spatial random process was explored in \cite{ChouQ:16,QueirozC:17}.  Both the GFT and KLT require solving an eigen-decomposition for each arrangement of points.  Point cloud attribute coding using a wavelet based transform, the region adaptive Haar transform (RAHT), was proposed and investigated in \cite{QueirozC:16,SandriRC:18}.  RAHT and a related lifting-based transform are used in PCC, as described in \cite{Schwarz:18}.

Our work in the present paper builds on both octree coding of geometry and RAHT coding of attributes.  For geometry processing, instead of using triangulations as geometric models within each leaf of the octree, we use an implicit surface defined by the wavelet coefficients of a volumetric B-spline of order $p=2$.  Compared to earlier approaches to geometry coding, our approach guarantees hole-free reconstruction of the surface, regardless of the level of quantization, due to the continuity properties of the B-spline.  For attribute processing, we show that RAHT has an interpretation as a volumetric B-spline of order $p=1$, and then we proceed to generalize it to B-splines of orders $p\geq2$.  Compared to RAHT ($p=1$), blocking artifacts with $p\geq2$ are eliminated, again due to continuity properties of the B-spline of orders $p\geq2$.

\section{Preliminaries}
\label{sec:preliminaries}

\subsection{Measure}

Our unique definition of measure is the foundation of our approach.

Let $\Omega$ be a set, and let $\sigma(\Omega)$ be a set of subsets of $\Omega$ such that $\sigma(\Omega)$ is a sigma-algebra, that is, closed under complementation and countably infinite unions.  A {\em measure} is a function $\mu:\sigma(\Omega)\rightarrow\mathbb{R}$ that assigns a real number to each set in $\sigma(\Omega)$ such that the measure of the union of any sequence of disjoint subsets is the sum of measures of the subsets.  Examples are the Lebesgue measure on the real line, the counting measure on the integers, and any probability measure on a probability space. 

We focus on the case where $\Omega=\mathbb{R}^3$ and $\sigma(\mathbb{R}^3)$ is the Borel sigma algebra of $\mathbb{R}^3$.  We suppose we are given a finite set of points in $\mathbb{R}^3$, say $\mathcal{X}=\{\bm{x}_1,\ldots,\bm{x}_N\}$.  For each set $\mathcal{M}\in\sigma(\mathbb{R}^3)$, we define $\mu(\mathcal{M})=|\mathcal{M}\cap \mathcal{X}|$ to be the number of such points in $\mathcal{M}$.  This is an example of a counting measure, albeit unconventional.

Let $f:\mathbb{R}^3\rightarrow\mathbb{R}$ be a real-valued function on $\mathbb{R}^3$.  The integral of $f$ over a set $\mathcal{M}\in\sigma(\mathbb{R}^3)$ with respect to measure $\mu$ is denoted $\int_\mathcal{M} f(\bm{x}) d\mu(\bm{x})$.  When $\mu$ is the counting measure with respect to $\mathcal{X}$, the integral is equal to
\begin{equation}
\int_\mathcal{M} f(\bm{x}) d\mu(\bm{x}) = \sum_{\bm{x}_n\in \mathcal{M}} f(\bm{x}_n).
\label{eqn:measure}
\end{equation}

\subsection{Hilbert Space}
\label{sec:hilbert}

A Hilbert space is a complete normed vector space equipped with an inner product that induces the norm.  Consider the Hilbert space $\mathcal{F}$ of real-valued functions $f:\mathbb{R}^3\rightarrow\mathbb{R}$ equipped with inner product
\begin{equation}
\left<f,g\right> = \int f(\bm{x})g(\bm{x}) d\mu(\bm{x})
= \sum_n f(\bm{x}_n)g(\bm{x}_n),
\label{eqn:innerproduct}
\end{equation}
where $f,g\in\mathcal{F}$, and $||f||\stackrel{\Delta}{=}\sqrt{\left<f,f\right>}$ is the induced norm.\footnote{Note that $||f||$ depends on the value of $f$ only at the points $\bm{x}_1,\ldots,\bm{x}_N$.  Hence strictly speaking, $||f||$ is only a pseudo-norm, as there are non-zero functions $f:\mathbb{R}^3\rightarrow\mathbb{R}$ such that $||f||=0$.  However, we call $||f||$ a norm with the usual understanding that $||f||=0$ implies $f=0$ a.e. (almost everywhere) with respect to measure $\mu$.  Alternatively, we consider the space of equivalence classes of functions, where two functions $f$ and $g$ are deemed equivalent if $f=g$ a.e.  Denoting by $\tilde f$ the equivalence class of functions equivalent to $f$, one can show that the set of equivalence classes $\tilde{\mathcal{F}}=\{\tilde f|f:\mathbb{R}^3\rightarrow\mathbb{R}\}$ is a vector space with a proper norm $||\tilde f||$ induced by the inner product $\left<\tilde f,\tilde g\right>$, which in turn is induced by the inner product $\left<f,g\right>$ between representatives.  The vector space $\tilde{\mathcal{F}}$ is clearly isomorphic to $\mathbb{R}^N$, and is hence a Hilbert space.  Thus we take $\mathcal{F}$ to be a Hilbert space with the usual ``almost everywhere'' understanding or with the equivalence class understanding \cite[p.~165]{Weir:73}.}

With the inner product and norm so determined, other properties of the Hilbert space follow:  Specifically, a vector $g\in\mathcal{F}$ is {\em orthogonal} to a vector $f\in\mathcal{F}$ iff $\left<f,g\right>=0$.  A vector $g\in\mathcal{F}$ is {\em orthogonal} to a subspace $\mathcal{F}_0\subseteq\mathcal{F}$ iff $g$ is orthogonal to all $f\in\mathcal{F}_0$.  A subspace $\mathcal{G}_0\subseteq\mathcal{F}$ is the {\em orthogonal complement} to a subspace $\mathcal{F}_0\subseteq\mathcal{F}$ iff for all $g\in\mathcal{G}_0$, $g$ is orthogonal to $\mathcal{F}_0$.
A point $f_0^*\in\mathcal{F}_0$ is the {\em projection} of a point $f\in\mathcal{F}$ onto the subspace $\mathcal{F}_0\in\mathcal{F}$ iff it minimizes $||f-f_0||$ over $f_0\in\mathcal{F}_0$.
The projection $f_0^*$ of $f$ onto $\mathcal{F}_0$, denoted $f\circ\mathcal{F}_0$, exists and is unique almost everywhere with respect to the measure $\mu$.
A necessary and sufficient condition for $f_0^*$ to be the projection of $f$ onto $\mathcal{F}_0$ is that the {\em approximation error} $(f-f_0^*)$ is orthogonal to $\mathcal{F}_0$ \cite[Thm.~2, p.~51]{Luenberger:69}.

\subsection{B\'ezier Volumes}

A {\em B\'ezier curve} of degree $m$ is a function on the unit interval $b:[0,1]\rightarrow\mathbb{R}$ specified as a linear combination of Bernstein polynomials, namely,
\begin{equation}
b(x)=\sum_{i=0}^m B_i b_{m,i}(x),
\end{equation}
where $B_0,\ldots,B_m$ are the coefficients of the linear combination, and
\begin{equation}
b_{m,i}(x)=\left(\begin{array}{c}m\\i\end{array}\right)x^i(1-x)^{m-i},
\end{equation}
$i=0,\ldots,m$, are the $m$th order Bernstein polynomials, which are polynomials of degree $m$ defined on the unit interval $[0,1]$.

Analogously, a {\em B\'ezier volume} (BV) of degree $m$ is a function on the unit cube $b:[0,1]^3\rightarrow\mathbb{R}$ specified as a linear combination of products of Bernstein polynomials, namely,
\begin{equation}
b(x,y,z)=\sum_{i=0}^m\sum_{j=0}^m\sum_{k=0}^m B_{ijk} b_{m,i}(x) b_{m,j}(y) b_{m,k}(z),
\end{equation}
where $B_{ijk}$, $i,j,k\in\{0,\ldots,m\}$, are the coefficients of the linear combination.

A function $b(x,y,z)$ is {\em tri-polynomial} of degree $m$ if it is a polynomial of degree $m$ in each of its coordinates when its other coordinates have any fixed value.  Thus a BV is tri-polynomial of degree $m$ over the unit cube.

\subsection{Cardinal B-splines}

\begin{figure}[!t]
\centering
\includegraphics[width=1.0\linewidth]{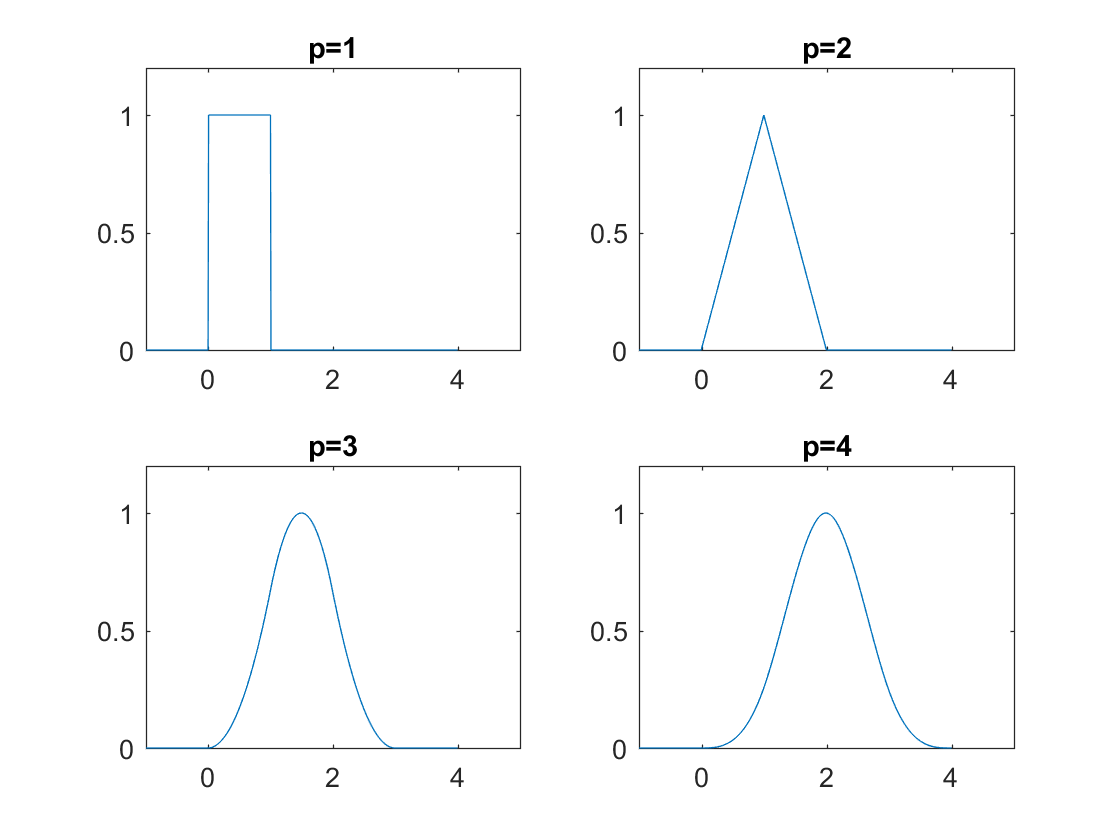}
\caption{Central B-spline basis functions of order $p$.}
\label{fig:bsplines}
\end{figure}

A {\em cardinal B-spline function} of order $p$ is a function on the real line $f:\mathbb{R}\rightarrow\mathbb{R}$ specified as a linear combination of B-spline basis functions of order $p$, namely
\begin{equation}
f(x)=\sum_{n\in\mathbb{Z}} F_n \phi^{(p)}(x-n),
\end{equation}
where $F_n$, $n\in\mathbb{Z}$, are the coefficients of the linear combination, and $\phi^{(p)}(x-n)$ is the B-spline basis function of order $p$ at integer shift $n$.  The B-spline basis function $\phi^{(p)}(x)$ can be defined for $p=1$ as
\begin{equation}
\phi^{(1)}(x)=\left\{\begin{array}{cc}1 & x\in[0,1] \\ 0 & \mbox{otherwise}\end{array}\right.
\end{equation}
and recursively for $p>1$ as
\begin{equation}
\phi^{(p)}(x)=\int \phi^{(1)}(t)\phi^{(p-1)}(x-t)dt
\end{equation}
for all $x$.  From this definition, it can be seen that $\phi^{(p)}(x)$ is the $p$-fold convolution of $\phi^{(1)}(x)$ with itself, and that the support of $\phi^{(p)}(x)$ is an interval of length $p$, as shown in Fig.~\ref{fig:bsplines}.

An alternative recursive definition for $p>1$ is
\begin{equation}
\phi^{(p)}(x)=\frac{1}{p-1}\left[x\phi^{(p-1)}(x)+(p-x)\phi^{(p-1)}(x-1)\right]\end{equation}
for $x\in[0,p]$ and $\phi^{(p)}(x)=0$ otherwise.
From this second definition it follows that $\phi^{(p)}(x)$ is a polynomial of degree $p-1$ on all integer shifts of the unit interval, $[n,n+1]$, $n\in\mathbb{Z}$, which we call {\em blocks}, and that $f(x)$ is piecewise polynomial of degree $p-1$ over each block.  It also follows that $f(x)$ is $C^{p-2}$ continuous, meaning that $f(x)$ and all of its derivatives up to its $(p-2)$th derivative are continuous, even at the breakpoints between blocks, which are called {\em knots}.

Analogously, a {\em cardinal B-spline volume} of order $p$ is a volumetric function $f:\mathbb{R}^3\rightarrow\mathbb{R}$ specified as a linear combination of vector integer shifts of a product of B-spline basis functions of order $p$, namely
\begin{equation}
f(\bm{x})=\sum_{\bm{n}\in\mathbb{Z}^3} F_{\bm{n}} \phi^{(p)}(\bm{x}-\bm{n}),
\label{eqn:bspline_expansion}
\end{equation}
where $F_{\bm{n}}$ is the coefficient of the linear combination at vector integer shift $\bm{n}\in\mathbb{Z}^3$, and $\phi^{(p)}(\bm{x})=\phi^{(p)}(x,y,z)=\phi^{(p)}(x)\phi^{(p)}(y)\phi^{(p)}(z)$ is the product of B-spline basis functions $\phi^{(p)}(x)$, $\phi^{(p)}(y)$, and $\phi^{(p)}(z)$.

It can be seen that $f(\bm{x})$ is tri-polynomial of degree $p-1$ over each shifted unit cube $[0,1]^3+\bm{n}$, or {\em block}.  Further, it can be seen that $f(\bm{x})$ is $C^{p-2}$ continuous between blocks.  Thus $f(\bm{x})$ can be considered to be a collection of B\'ezier volumes of order $p-1$ linked together such that the overall function is $C^{p-2}$ continuous.

\subsection{Approximation}
\label{sec:approximation}

Let $\phi_{0,\bm{0}}(\bm{x})=\phi^{(p)}(\bm{x}-\bm{n}_0)$ be the {\em central} cardinal B-spline basis function, that is, the cardinal B-spline basis function centered on the origin $\bm{0}$ (if $p$ is even) or on the unit cube $[0,1]^3$ (if $p$ is odd).  Let $\phi_{0,\bm{n}}(\bm{x})=\phi_{0,\bm{0}}(\bm{x}-\bm{n})$ be this central cardinal B-spline basis function shifted by integer vector $\bm{n}$.

For notational simplicity, we here suppress the dependence of $\phi_{0,\bm{n}}$ on $p$.  We discuss specific values of $p$ in Section~\ref{sec:attributes}.

Let $\mathcal{F}$ be the Hilbert space of all functions $f:\mathbb{R}^3\rightarrow\mathbb{R}$ under the inner product (\ref{eqn:innerproduct}), as defined in Subsection~\ref{sec:hilbert}.  Define the subspace $\mathcal{F}_0\subseteq\mathcal{F}$ as
\begin{equation}
\mathcal{F}_0=\left\{f_0\in\mathcal{F}\left|\exists\{F_{\bm{n}}\} \mbox{\;s.t.\;} f_0(\bm{x})=\sum_{\bm{n}\in\mathbb{Z}^3} F_{\bm{n}} \phi_{0,\bm{n}}(\bm{x})\right.\right\}.
\end{equation}
This is the subspace of all functions that are tri-polynomial of degree $p-1$ over the blocks $\{[0,1]^3+\bm{n}|\bm{n}\in\mathbb{Z}^3\}$.

Any function $f\in\mathcal{F}$ can be approximated by a function $f_0^*\in\mathcal{F}_0$, where $f_0^*$ is the projection of $f$ onto $\mathcal{F}_0$, denoted $f_0^*=f\circ\mathcal{F}_0$.  Let $\{F_{\bm{n}}^*\}$ be coefficients such that $f_0^*(\bm{x})=\sum_{\bm{n}\in\mathcal{Z}^3}F_{\bm{n}}^*\phi_{0,\bm{n}}(\bm{x})$.  Under the inner product (\ref{eqn:innerproduct}), the squared error between $f$ and $f_0^*$,
\begin{equation}
||f-f_0^*||^2 = \sum_{i=1}^N(f(\bm{x}_i)-f_0^*(\bm{x}_i))^2,
\end{equation}
depends on the values of $f_0^*$ only at the points $\bm{x}_1,\ldots,\bm{x}_N$, which in turn depend on any particular coefficient $F_{\bm{n}}^*$ only if $\phi_{0,\bm{n}}(\bm{x}_i)\neq 0$ for some $\bm{x}_i$, $i=1\ldots,N$.  Let
\begin{equation}
\mathcal{N}_0 = \left\{\bm{n}\left|\exists i\in\{1,\ldots,N\}\;\mbox{s.t.}\;\phi_{0,\bm{n}}(\bm{x}_i)\neq 0\right.\right\}
\end{equation}
be the set of vector integer shifts $\bm{n}$ such that $\phi_{0,\bm{n}}(\bm{x}_i)\neq 0$ for some $\bm{x}_i$.  This set is {\em finite} because $\phi_{0,\bm{0}}$ has bounded support, and any of its shifts far away from the points $\bm{x}_1,\ldots,\bm{x}_N$ will not include any such point in its support.  Let $\{\bm{n}_i\}$ denote the finite set of shifts in $\mathcal{N}_0$.

For any $\bm{n}\not\in\mathcal{N}_0$, assign $F_{\bm{n}}^*=0$, and for any $\bm{n}\in\mathcal{N}_0$, solve for $F_{\bm{n}}^*$ by noting that the approximation error $(f-f_0^*)$ must be orthogonal to $\phi_{0,\bm{n}}$ for all $\bm{n}\in\mathbb{Z}^3$.  In particular, for all $\bm{n}_i\in\mathcal{N}_0$,
\begin{equation}
0 = \left<\phi_{0,\bm{n}_i},f-f_0^*\right>
  = \left<\phi_{0,\bm{n}_i},f\right> - \left<\phi_{0,\bm{n}_i},f_0^*\right>,
\end{equation}
or
\begin{equation}
\left<\phi_{0,\bm{n}_i},f\right>
= \left<\phi_{0,\bm{n}_i},f_0^*\right> 
= \sum_{\bm{n}_j\in\mathcal{N}_0} \left<\phi_{0,\bm{n}_i},\phi_{0,\bm{n}_j}\right> F_{\bm{n}_j}^*.
\end{equation}
These are the {\em normal equations}.  In vector form,
\begin{equation}
\Phi_0^Tf = \Phi_0^T\Phi_0 F^*
\end{equation}
where $\Phi_0^Tf$ is shorthand for the $|\mathcal{N}_0|\times1$ vector $\left[\left<\phi_{0,\bm{n}_i},f\right>\right]$, $\Phi_0^T\Phi_0$ is shorthand for the $|\mathcal{N}_0|\times|\mathcal{N}_0|$ matrix $\left[\left<\phi_{0,\bm{n}_i},\phi_{0,\bm{n}_j}\right>\right]$, and $F^*$ is the $|\mathcal{N}_0|\times1$  vector $[F_{\bm{n}_i}^*]$.  If $\Phi_0^T\Phi_0$ is invertible, then one may solve for $F^*$ explicitly as
\begin{equation}
F^* = (\Phi_0^T\Phi_0)^{-1}\Phi_0^Tf.
\label{eqn:lowpass}
\end{equation}

\subsection{Multiresolution Approximation}
\label{sec:multiresolution}




To obtain approximations at different resolutions, the cardinal B-spline basis functions can be scaled by a factor of $2^{-\ell}$, where $\ell$ is the {\em scale} or {\em level of detail} or simply {\em level}.  To be specific, define
\begin{equation}
\phi_{\ell,\bm{n}}(\bm{x})=\phi_{0,\bm{n}}(2^\ell\bm{x})
\end{equation}
as the cardinal B-spline basis function at level $\ell$ and shift $\bm{n}$.
Define the subspace $\mathcal{F}_\ell\subseteq\mathcal{F}$ as
\begin{equation}
\mathcal{F}_\ell=\left\{f_\ell\in\mathcal{F}\left|\exists\{F_{\ell,\bm{n}}\} \mbox{\;s.t.\;} f_\ell(\bm{x})=\sum_{\bm{n}\in\mathbb{Z}^3} F_{\ell,\bm{n}} \phi_{\ell,\bm{n}}(\bm{x})\right.\right\}.
\label{eqn:F_ell}
\end{equation}
This is the subspace of all functions that are tri-polynomial of degree $p-1$ over the blocks at level $\ell$, $\{2^{-\ell}([0,1]^3+\bm{n})|\bm{n}\in\mathbb{Z}^3\}$.  Since the blocks at level $\ell$ are refined by the blocks at level $\ell+1$, it is clear that if a function $f_\ell$ is tri-polynomial over the blocks at level $\ell$, i.e., $f_\ell\in\mathcal{F}_\ell$, then it is also tri-polynomial over the blocks at level $\ell+1$, i.e., $f_\ell\in\mathcal{F}_{\ell+1}$.  Hence $\mathcal{F}_\ell\subseteq\mathcal{F}_{\ell+1}$ and
\begin{equation}
\mathcal{F}_0\subseteq\mathcal{F}_1\subseteq\ldots\subseteq\mathcal{F}_\ell\subseteq\mathcal{F}_{\ell+1}\subseteq\ldots\subseteq\mathcal{F}
\label{eqn:nested}
\end{equation}
is a nested sequence of subspaces whose resolution increases with $\ell$.

Let $f_\ell^*=f\circ\mathcal{F}_\ell$ and $f_{\ell+1}^*=f\circ\mathcal{F}_{\ell+1}$ be the projections of $f$ onto $\mathcal{F}_\ell$ and $\mathcal{F}_{\ell+1}$, respectively.  Then by the Pythagorean theorem \cite[Sec.~3.3]{Luenberger:69}, for all $f_\ell\in\mathcal{F}_\ell\subseteq\mathcal{F}_{\ell+1}$,
\begin{equation}
||f-f_\ell||^2=||f-f_{\ell+1}^*||^2+||f_{\ell+1}^*-f_\ell||^2.
\label{eqn:pythagorean}
\end{equation}
Then since $f_\ell^*=f\circ\mathcal{F}_\ell$ minimizes $||f-f_\ell||^2$ over all $f_\ell\in\mathcal{F}_\ell$, by (\ref{eqn:pythagorean}) $f_\ell^*$ must also minimize $||f_{\ell+1}^*-f_\ell||^2$ over all $f_\ell\in\mathcal{F}_\ell$, and hence $f_\ell^*=f_{\ell+1}^*\circ\mathcal{F}_\ell$.  That is, projecting $f$ onto $\mathcal{F}_\ell$ can be done in two steps, by first projecting onto $\mathcal{F}_{\ell+1}$ (i.e., $f_{\ell+1}^*=f\circ\mathcal{F}_{\ell+1}$) and then onto $\mathcal{F}_\ell$ (i.e., $f_\ell^*=f_{\ell+1}^*\circ\mathcal{F}_\ell$).  Alternatively, $f\circ\mathcal{F}_\ell=f\circ\mathcal{F}_{\ell+1}\circ\mathcal{F}_\ell$.

Paralleling the development in the previous subsection, let
\begin{equation}
\mathcal{N}_\ell = \left\{\bm{n}\left|\exists i\in\{1,\ldots,N\}\;\mbox{s.t.}\;\phi_{\ell,\bm{n}}(\bm{x}_i)\neq 0\right.\right\}
\label{eqn:N_ell}
\end{equation}
be the finite set of vector integer shifts $\bm{n}$ such that $\phi_{\ell,\bm{n}}(\bm{x}_i)\neq 0$ for some $\bm{x}_i$. Then for all $\bm{n}_i\in\mathcal{N}_\ell$,
\begin{equation}
0 = \left<\phi_{\ell,\bm{n}_i},f-f_\ell^*\right>
  = \left<\phi_{\ell,\bm{n}_i},f\right> - \left<\phi_{\ell,\bm{n}_i},f_\ell^*\right>,
\end{equation}
or
\begin{equation}
\left<\phi_{\ell,\bm{n}_i},f\right>
= \left<\phi_{\ell,\bm{n}_i},f_\ell^*\right> 
= \sum_{\bm{n}_j\in\mathcal{N}_\ell} \left<\phi_{\ell,\bm{n}_i},\phi_{\ell,\bm{n}_j}\right> F_{\ell,\bm{n}_j}^*,
\label{eqn:phi_ell_T_f_i}
\end{equation}
where $f_\ell^*=\sum_{\bm{n}_j\in\mathcal{N}_\ell}F_{\ell,\bm{n}_j}^*\phi_{\ell,\bm{n}_j}$.  In vector form, (\ref{eqn:phi_ell_T_f_i}) can be expressed
\begin{equation}
\Phi_\ell^Tf = \Phi_\ell^T\Phi_\ell F_\ell^*
\end{equation}
where $\Phi_\ell^Tf$ is shorthand for the $|\mathcal{N}_\ell|\times1$ vector $[\left<\phi_{\ell,\bm{n}_i},f\right>]$, $\Phi_\ell^T\Phi_\ell$ is shorthand for the $|\mathcal{N}_\ell|\times|\mathcal{N}_\ell|$ matrix $[\left<\phi_{\ell,\bm{n}_i},\phi_{\ell,\bm{n}_j}\right>]$, and $F_\ell^*$ is the $|\mathcal{N}_\ell|\times1$  vector $[F_{\ell,\bm{n}_j}^*]$.  If $\Phi_\ell^T\Phi_\ell$ is invertible, then one may solve for $F_\ell^*$ explicitly as
\begin{equation}
F_\ell^* = (\Phi_\ell^T\Phi_\ell)^{-1}\Phi_\ell^Tf.
\label{eqn:F_ell_star}
\end{equation}
In turn, one may compute $\Phi_\ell^Tf$ and $\Phi_\ell^T\Phi_\ell$ recursively from $\Phi_{\ell+1}^Tf$ and $\Phi_{\ell+1}^T\Phi_{\ell+1}$, respectively, as follows.

Since $\phi_{\ell,\bm{n}}\in\mathcal{F}_\ell\subseteq\mathcal{F}_{\ell+1}$, there exist coefficients $\{a_{\bm{k}}\}$ not depending on $\ell$ such that
\begin{equation}
\phi_{\ell,\bm{n}}
= \sum_{\bm{k}\in\mathbb{Z}^3} a_{\bm{k}} \phi_{\ell+1,2\bm{n}+\bm{k}}
= \sum_{\bm{k}'\in\mathbb{Z}^3} a_{\bm{k}'-2\bm{n}} \phi_{\ell+1,\bm{k}'}
\label{eqn:twoscale}
\end{equation}
Equation (\ref{eqn:twoscale}) is known as the {\em two-scale equation}.  From this equation, it follows that
\begin{equation}
\left<\phi_{\ell,\bm{n}_i},f\right> = \sum_{\bm{n}_j\in\mathcal{N}_{\ell+1}} a_{\bm{n}_j-2\bm{n}_i} \left<\phi_{\ell+1,\bm{n}_j},f\right>
\end{equation}
and
\begin{eqnarray}
\lefteqn{\left<\phi_{\ell,\bm{n}_i},\phi_{\ell,\bm{n}_j}\right>} \\
& \displaystyle = \sum_{\bm{n}_k\in\mathcal{N}_{\ell+1}} \sum_{\bm{n}_l\in\mathcal{N}_{\ell+1}} a_{\bm{n}_k-2\bm{n}_i} \left<\phi_{\ell+1,\bm{n}_k},\phi_{\ell+1,\bm{n}_l}\right> a_{\bm{n}_l-2\bm{n}_j}. \nonumber 
\end{eqnarray}
In vector form,
\begin{equation}
\Phi_\ell^Tf = A_\ell\Phi_{\ell+1}^Tf
\label{eqn:phi_ell_T_f}
\end{equation}
and
\begin{equation}
\Phi_\ell^T\Phi_\ell = A_\ell\Phi_{\ell+1}^T\Phi_{\ell+1}A_\ell^T,
\label{eqn:phi_ell_T_phi_ell}
\end{equation}
where $A_\ell=[a_{\bm{n}_j-2\bm{n}_i}]$.
Another useful recursion is
\begin{eqnarray}
\left<\phi_{\ell,\bm{n}_i},\phi_{\ell+1,\bm{n}_j}\right>
= \sum_{\bm{n}_k\in\mathcal{N}_{\ell+1}} a_{\bm{n}_k-2\bm{n}_i} \left<\phi_{\ell+1,\bm{n}_k},\phi_{\ell+1,\bm{n}_j}\right>,
\end{eqnarray}
or
\begin{equation}
\Phi_\ell^T\Phi_{\ell+1} = A_\ell\Phi_{\ell+1}^T\Phi_{\ell+1}
\label{eqn:phi_ell_T_phi_ell1}
\end{equation}
in vector form.

\subsection{Wavelets}
\label{sec:wavelets}

Let $\mathcal{G}_\ell$ be the orthogonal complement of $\mathcal{F}_\ell$ in $\mathcal{F}_{\ell+1}$, i.e.,
\begin{equation}
\mathcal{F}_{\ell+1} = \mathcal{F}_\ell\oplus\mathcal{G}_\ell.
\end{equation}
Applying this recursively, we have
\begin{equation}
\mathcal{F}_{\ell+1} = \mathcal{F}_0\oplus\mathcal{G}_0\oplus\mathcal{G}_1\oplus\cdots\oplus\mathcal{G}_\ell,
\end{equation}
so that any function $f_{\ell+1}\in\mathcal{F}_{\ell+1}$ can be written as the sum of orthogonal functions,
\begin{equation}
f_{\ell+1} = f_0 + g_0 + g_1 + \cdots + g_\ell.
\end{equation}
The coefficients of $f_0$ in the basis for $\mathcal{F}_0$ are {\em low pass} coefficients, while the coefficients of $g_\ell$ in the basis for $\mathcal{G}_\ell$ are {\em high pass} or {\em wavelet} coefficients.  The function $f$ can be communicated by quantizing and entropy coding its low pass and wavelet coefficients.  This is efficient because most of the energy in $f$ is in its low pass coefficients and its low-level wavelet coefficients.

To compute the low pass coefficients, (\ref{eqn:lowpass}) can be used, while to compute the wavelet coefficients, we first need to establish a basis for each $\mathcal{G}_\ell$.

First, some definitions:  For each $\ell$, let $\Phi_\ell=[\phi_{\ell,\bm{n}_j}]$ be a row vector containing the functions $\phi_{\ell,\bm{n}_j}$, $\bm{n}_j\in\mathcal{N}_\ell$, and let $F_\ell$ be a column vector of $|\mathcal{N}_\ell|$ coefficients.  Let $\Phi_\ell F_\ell$ denote a function $f_\ell=\sum_{\bm{n}_j\in\mathcal{N}_\ell}F_{\ell,\bm{n}_j}\phi_{\ell,\bm{n}_j}$ in $\mathcal{F}_{\ell+1}$.  Let $\Phi_\ell^Tf=[\left<\phi_{\ell,\bm{n}_i},f\right>]$ denote the  column vector of inner products of the functions $\phi_{\ell,\bm{n}_i}$, $\bm{n}_i\in\mathcal{N}_\ell$, with the function $f$.  Similarly, let $\Phi_\ell^T[f_1,\ldots,f_n]=[\left<\phi_{\ell,\bm{n}_i},f_j\right>]$ denote the matrix of inner products of the functions $\phi_{\ell,\bm{n}_i}$, $\bm{n}_i\in\mathcal{N}_\ell$, with the functions $f_j$ $j=1,\ldots,n$.

Consider now the subspace $\mathcal{G}_\ell\subseteq\mathcal{F}_{\ell+1}$ defined by
\begin{equation}
\mathcal{G}_\ell=\left\{\Phi_{\ell+1}F_{\ell+1}\left|\Phi_\ell^T\Phi_{\ell+1}F_{\ell+1}=0\right.\right\}.
\label{eqn:Gell}
\end{equation}
Clearly, $\mathcal{G}_\ell$ is a subspace of $\mathcal{F}_{\ell+1}$ and is orthogonal to $\mathcal{F}_\ell=\{\Phi_\ell F_\ell\}$, and hence $\mathcal{G}_\ell$ is the orthogonal complement of $\mathcal{F}_\ell$ in $\mathcal{F}_{\ell+1}$.
If $N_{\ell+1}$ is the dimension of $\mathcal{F}_{\ell+1}$ and $N_\ell$ is the dimension of $\mathcal{F}_\ell$, then $N_{\ell+1}-N_\ell$ is the dimension of $\mathcal{G}_\ell$.  Typically, $N_\ell=|\mathcal{N}_\ell|$.  (When $\ell$ is large, the dimension $N_\ell$ of $\mathcal{F}_\ell$ may be lower than $|\mathcal{N}_\ell|$, due to the finite number of points $\bm{x}_1,\ldots,\bm{x}_N$.  We will point out what to do about this case later in this subsection.)

One way to construct an explicit basis for $\mathcal{G}_\ell$ is as follows.  Partition the $N_\ell\times N_{\ell+1}$ matrix $\Phi_\ell^T\Phi_{\ell+1}$ into an $N_\ell\times N_\ell$ matrix $A$ and an $N_\ell\times(N_{\ell+1}-N_\ell)$ matrix $B$, as $\Phi_\ell^T\Phi_{\ell+1}=[A\;B]$.  Similarly partition the $N_{\ell+1}$ dimensional vector $F_{\ell+1}$ into an $N_\ell$ dimensional vector $F^a$ and an $N_{\ell+1}-N_\ell$ dimensional vector $F^b$, as
\begin{equation}
F_{\ell+1}=\left[\begin{array}{c}F^a\\F^b\end{array}\right].
\label{eqn:polyphase}
\end{equation}
Then for all $F_{\ell+1}$ satisfying $\Phi_\ell^T\Phi_{\ell+1}F_{\ell+1}=0$ in (\ref{eqn:Gell}), we have
\begin{equation}
[A\;B]\left[\begin{array}{c}F^a\\F^b\end{array}\right] = 0
\end{equation}
\begin{equation}
A^{-1}[A\;B]\left[\begin{array}{c}F^a\\F^b\end{array}\right] = 0
\end{equation}
\begin{equation}
[I^a\;A^{-1}B]\left[\begin{array}{c}F^a\\F^b\end{array}\right] = 0
\end{equation}
where $I^a$ is the $N_\ell\times N_\ell$ identity matrix.  Hence $F^a=-A^{-1}BF^b$, and therefore $F_{\ell+1}=ZF^b$, where
\begin{equation}
Z=\left[\begin{array}{c}-A^{-1}B\\I^b\end{array}\right]
\label{eqn:nullspace}
\end{equation}
is an $N_{\ell+1}\times(N_{\ell+1}-N_\ell)$ matrix containing the $(N_{\ell+1}-N_\ell)\times(N_{\ell+1}-N_\ell)$ identity matrix $I^b$.  Thus
\begin{equation}
\mathcal{G}_\ell=\left\{\left.\Phi_{\ell+1}ZF^b\right |F^b\in\mathbb{R}^{N_{\ell+1}-N_\ell}\right\},
\end{equation}
and the $(N_{\ell+1}-N_\ell)$ functions in the row vector
\begin{equation}
\Psi_\ell = \Phi_{\ell+1}Z
\end{equation}
form an explicit basis for $\mathcal{G}_\ell$.

More generally, instead of (\ref{eqn:nullspace}), any matrix $Z$ whose columns span the null space of $\Phi_\ell^T\Phi_{\ell+1}$ may be used to form a basis for $\mathcal{G}_\ell$.

If the rank $N_\ell$ of $\Phi_\ell^T\Phi_{\ell+1}$ is less than $|\mathcal{N}_\ell|$, then indices $\bm{n}_i\in\mathcal{N}_\ell$ (corresponding to rows of $\Phi_\ell^T\Phi_{\ell+1}$) may be removed until $|\mathcal{N}_\ell|$ is equal to the rank $N_\ell$ of $\Phi_\ell^T\Phi_{\ell+1}$, i.e., until $\Phi_\ell^T\Phi_{\ell+1}$ has full rank.  As will be seen later, this will ensure a critically sampled transform.

Now that we have established a basis for $\mathcal{G}_\ell$, any function $f_{\ell+1}=\Phi_{\ell+1}F_{\ell+1}\in\mathcal{F}_{\ell+1}$ can be decomposed as the sum of functions $f_\ell=\Phi_\ell F_\ell\in\mathcal{F}_\ell$ and $g_\ell=\Psi_\ell G_\ell\in\mathcal{G}_\ell$, specifically,
\begin{equation}
\Phi_{\ell+1}F_{\ell+1} = \left[\begin{array}{cc}\Phi_\ell & \Psi_\ell\end{array}\right]
\left[\begin{array}{c}F_\ell \\ G_\ell\end{array}\right].
\label{eqn:deomposition}
\end{equation}

How to obtain $F_\ell$ and $G_\ell$ from $F_{\ell+1}$, and vice-versa, is the subject of the next subsection.

\subsection{Two-Channel Filter Banks}
\label{sec:filterbanks}

Formulas for an analysis filter bank (which produce coefficients $F_\ell$ and $G_\ell$ from coefficients $F_{\ell+1}$) and a synthesis filter bank (which produce coefficients $F_{\ell+1}$ from coefficients $F_\ell$ and $G_\ell$) may be obtained by taking the inner product of (\ref{eqn:deomposition}) with the $N_\ell$ functions in $\Phi_\ell$ and the $N_{\ell+1}-N_\ell$ functions in $\Psi_\ell$,
\begin{equation}
\left[\begin{array}{cc}\Phi_\ell & \Psi_\ell\end{array}\right]^T\Phi_{\ell+1}F_{\ell+1} = \left[\begin{array}{cc}\Phi_\ell^T\Phi_\ell & 0 \\ 0 & \Psi_\ell^T\Psi_\ell \end{array}\right]
\left[\begin{array}{c}F_\ell \\ G_\ell\end{array}\right],
\end{equation}
yielding
\begin{equation}
\left[\begin{array}{c}F_\ell \\ G_\ell\end{array}\right] = \left[\begin{array}{cc}\Phi_\ell^T\Phi_\ell & 0 \\ 0 & \Psi_\ell^T\Psi_\ell \end{array}\right]^{-1}\left[\begin{array}{c}\Phi_\ell^T\Phi_{\ell+1} \\ \Psi_\ell^T\Phi_{\ell+1}\end{array}\right]F_{\ell+1}
\label{eqn:analysis}
\end{equation}
for the analysis and
\begin{equation}
F_{\ell+1} = \left[\begin{array}{c}\Phi_\ell^T\Phi_{\ell+1} \\ \Psi_\ell^T\Phi_{\ell+1}\end{array}\right]^{-1}\left[\begin{array}{cc}\Phi_\ell^T\Phi_\ell & 0 \\ 0 & \Psi_\ell^T\Psi_\ell \end{array}\right]
\left[\begin{array}{c}F_\ell \\ G_\ell\end{array}\right]
\label{eqn:synthesis}
\end{equation}
for the synthesis.

The elements of $F_{\ell+1}$ may be re-ordered such that the first $N_\ell$ elements $F^a$ are designated {\em even} and the last $N_{\ell+1}-N_\ell$ elements $F^b$ are designated {\em odd}, as in (\ref{eqn:polyphase}), analogous to a polyphase decomposition.  Correspondingly, the elements of $\Phi_{\ell+1}$ may be re-ordered as $\Phi_{\ell+1}=\left[\begin{array}{cc}\Phi_{\ell+1}^a & \Phi_{\ell+1}^b\end{array}\right]$. Then the analysis and synthesis (\ref{eqn:analysis}) and (\ref{eqn:synthesis}) may be re-written
\begin{equation}
\left[\begin{array}{c}F_\ell \\ G_\ell\end{array}\right]
= T_\ell
\left[\begin{array}{c}F_{\ell+1}^a \\ F_{\ell+1}^b\end{array}\right], \;
\left[\begin{array}{c}F_{\ell+1}^a \\ F_{\ell+1}^b\end{array}\right]
= T_\ell^{-1}
\left[\begin{array}{c}F_\ell \\ G_\ell\end{array}\right],
\end{equation}
where
\begin{eqnarray}
T_\ell
& = & \left[\begin{array}{cc}
(\Phi_\ell^T\Phi_\ell)^{-1}\Phi_\ell^T\Phi_{\ell+1}^a &
(\Phi_\ell^T\Phi_\ell)^{-1}\Phi_\ell^T\Phi_{\ell+1}^b \\
(\Psi_\ell^T\Psi_\ell)^{-1}\Psi_\ell^T\Phi_{\ell+1}^a &
(\Psi_\ell^T\Psi_\ell)^{-1}\Psi_\ell^T\Phi_{\ell+1}^b
\end{array}\right] \\
& = & \left[\begin{array}{cc}A & B \\ C & D\end{array}\right],
\label{eqn:ABCD}
\end{eqnarray}
where $A$, $B$, $C$, and $D$ are the blocks of the matrix, and (using either of the block matrix inverse formulas)
\begin{widetext}
\begin{eqnarray}
T_\ell^{-1}
= \left[\begin{array}{cc}A & B \\ C & D\end{array}\right]^{-1}
& = & \left[\begin{array}{cc}
A^{-1}+A^{-1}B(D-CA^{-1}B)^{-1}CA^{-1} &
-A^{-1}B(D-CA^{-1}B)^{-1} \\
-(D-CA^{-1}B)^{-1}CA^{-1} &
(D-CA^{-1}B)^{-1}CA^{-1}\end{array}\right]
\label{eqn:ABCDinv1} \\
& = & \left[\begin{array}{cc}
(A-BD^{-1}C)^{-1} &
-(A-BD^{-1}C)^{-1}BD^{-1} \\
-D^{-1}C(A-BD^{-1}C)^{-1} &
D^{-1}+D^{-1}C(A-BD^{-1}C)^{-1}BD^{-1}\end{array}\right].
\label{eqn:ABCDinv2}
\end{eqnarray}
\end{widetext}

\begin{figure}[!t]
\centering
\includegraphics[width=0.65\columnwidth]{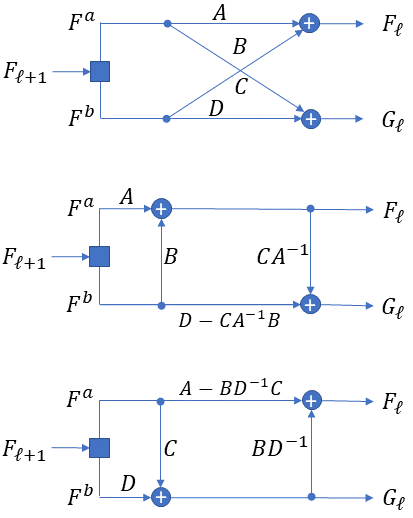}
\caption{Three equivalent analysis filter banks, from top: (a) Butterfly, (b) Lifting with prediction followed by update, and (c) Lifting with update followed by prediction.}
\label{fig:analysis}
\end{figure}

\begin{figure}[!t]
\centering
\includegraphics[width=0.65\columnwidth]{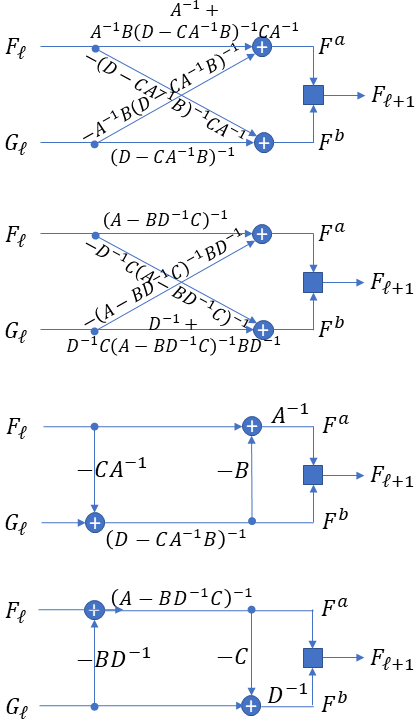}
\caption{Four equivalent synthesis filter banks, from top: (a) Butterfly using (\ref{eqn:ABCDinv1}), (b) Butterfly using (\ref{eqn:ABCDinv2}), (c) Lifting with prediction followed by update, and (d) Lifting with update followed by prediction.}
\label{fig:synthesis}
\end{figure}

The block matrix expressions for $T_\ell$ and $T_\ell^{-1}$ lead to three equivalent analysis filter banks, shown in Fig.~\ref{fig:analysis}, and four equivalent synthesis filter banks, shown in Fig.~\ref{fig:synthesis}.  The filter banks may be nested, recursively decomposing the low pass coefficients.

\subsection{Orthogonality}
\label{sec:orthgonality}

From the above construction, the elements of $\Phi_\ell$ are orthogonal to the elements of $\Psi_\ell$, but the elements of $\Phi_\ell$ are not necessarily orthogonal to each other, and the elements of $\Psi_\ell$ are not necessarily orthogonal to each other.  However, if desired, they may be orthogonalized (and normalized) by finding an $N_\ell\times N_\ell$ matrix $R_\ell$ and an $(N_{\ell+1}-N_\ell)\times(N_{\ell+1}-N_\ell)$ matrix $S_\ell$ such that the elements of the row vector $\Phi_\ell R_\ell$ are orthonormal (i.e., $R_\ell^T\Phi_\ell^T\Phi_\ell R_\ell=I$) and the elements of the row vector $\Psi_\ell S_\ell$ are orthonormal (i.e., $S_\ell^T\Psi_\ell^T\Psi_\ell S_\ell=I$), respectively.  To find $R_\ell$, perform the eigen-decomposition of the positive definite matrix $\Phi_\ell^T\Phi_\ell=U\Lambda U^T$, where $U$ is the matrix whose columns are orthonormal eigenvectors and $\Lambda$ is the diagonal matrix of positive eigenvalues.  Then $R_\ell=U\Lambda^{-1/2}$ since in that case $R_\ell^T\Phi_\ell^T\Phi_\ell R_\ell=R_\ell^TU\Lambda U^TR_\ell=I$.  Similarly,  $S_\ell=V\Delta^{-1/2}$, where $V$ is the matrix of eigenvectors of $\Psi_\ell^T\Psi_\ell$ and $\Delta$ is the diagonal matrix of positive eigenvalues.

Denoting the row vectors of orthonormal basis functions for $\mathcal{F}_\ell$ and $\mathcal{G}_\ell$ by
\begin{equation}
\bar{\Phi}_\ell=\Phi_\ell R_\ell, \;
\bar{\Psi}_\ell=\Psi_\ell S_\ell,
\end{equation}
and denoting their coefficients by
\begin{equation}
\bar{F}_\ell=R_\ell^{-1}F_\ell, \;
\bar{G}_\ell=S_\ell^{-1}G_\ell,
\end{equation}
orthonormal versions of the analysis and synthesis may be written
\begin{eqnarray}
\left[\begin{array}{c}\bar{F}_\ell \\ \bar{G}_\ell\end{array}\right]
& = & \left[\begin{array}{c}
R_\ell^{-1}(\Phi_\ell^T\Phi_\ell)^{-1}\Phi_\ell^T\Phi_{\ell+1} \\
S_\ell^{-1}(\Psi_\ell^T\Psi_\ell)^{-1}\Psi_\ell^T\Phi_{\ell+1}
\end{array}\right]F_{\ell+1} \\
& = & \left[\begin{array}{c}
(R_\ell^T\Phi_\ell^T\Phi_\ell R_\ell)^{-1}R_\ell^T\Phi_\ell^T\Phi_{\ell+1} \\
(S_\ell^T\Psi_\ell^T\Psi_\ell S_\ell)^{-1}S_\ell^T\Psi_\ell^T\Phi_{\ell+1}
\end{array}\right]F_{\ell+1} \\
& = & \left[\begin{array}{c}
R_\ell^T\Phi_\ell^T\Phi_{\ell+1}R_{\ell+1} \\
S_\ell^T\Psi_\ell^T\Phi_{\ell+1}S_{\ell+1}
\end{array}\right]\bar{F}_{\ell+1} \\
& = & \left[\begin{array}{c}
\bar{\Phi}_\ell^T\bar{\Phi}_{\ell+1} \\
\bar{\Psi}_\ell^T\bar{\Phi}_{\ell+1}
\end{array}\right]\bar{F}_{\ell+1}
= \bar{T}_\ell\bar{F}_{\ell+1},
\end{eqnarray}
where the forward transform $\bar{T}_\ell$ is orthonormal, and hence its inverse is simply its transpose, $\bar{T}_\ell^{-1} = \bar{T}_\ell^T$.  Thus
\begin{equation}
\bar{F}_{\ell+1} = \left[\begin{array}{c}
\bar{\Phi}_\ell^T\bar{\Phi}_{\ell+1} \\
\bar{\Psi}_\ell^T\bar{\Phi}_{\ell+1}
\end{array}\right]^T
\left[\begin{array}{c}\bar{F}_\ell \\ \bar{G}_\ell\end{array}\right]
= \bar{T}_\ell^T
\left[\begin{array}{c}\bar{F}_\ell \\ \bar{G}_\ell\end{array}\right].
\end{equation}
Like $T_\ell$, $\bar{T}_\ell$ can also be written as a block matrix,
\begin{equation}
\bar{T}_\ell
= \left[\begin{array}{cc}
\bar{\Phi}_\ell^T\bar{\Phi}_{\ell+1}^a &
\bar{\Phi}_\ell^T\bar{\Phi}_{\ell+1}^b \\
\bar{\Psi}_\ell^T\bar{\Phi}_{\ell+1}^a &
\bar{\Psi}_\ell^T\bar{\Phi}_{\ell+1}^b
\end{array}\right]
= \left[\begin{array}{cc}
\bar A &
\bar B \\
\bar C &
\bar D
\end{array}\right],
\end{equation}
and hence $\bar{T}_\ell$ and $\bar{T}_\ell^{-1}$ have the same structures shown in Figs.~\ref{fig:analysis} and~\ref{fig:synthesis}, except that the expressions for the synthesis butterfly are much simpler: $\bar A^T$, $\bar C^T$, $\bar B^T$, and $\bar D^T$.

Note that because the transforms are orthonormal at each stage of the wavelet transform, the overall transform is orthonormal as well as critically sampled.

\section{Attribute Coding using Region Adaptive Hierarchical Transforms}
\label{sec:attributes}

In this section, we show how to represent and compress the real-valued attributes of a point cloud using volumetric functions.  We assume that we are given, at the encoder, a set of point locations $\bm{x}_1,\ldots,\bm{x}_N$ and a set of corresponding attributes $f_1,\ldots,f_N$.  We also assume that the point locations can be communicated to the decoder without loss.  The problem of attribute compression is to reproduce approximate attributes, $\hat f_1,\ldots,\hat f_N$, at the decoder, subject to a constraint on the number of bits communicated, given the point locations as side information.  We assume for simplicity that the attributes are scalar.  Vector attributes can be treated component-wise.

Our approach is the following.  At the encoder, a volumetric B-spline of order $p$ is fit to the values $f_1,\ldots,f_N$ at locations $\bm{x}_1,\ldots,\bm{x}_N$, and its wavelet coefficients are quantized and entropy coded.  At the decoder, the wavelet coefficients are entropy decoded and dequantized, and the volumetric B-spline is reconstructed.  Finally the reconstructed volumetric B-spline is sampled at the locations $\bm{x}_1,\ldots,\bm{x}_N$, and the corresponding values $\hat f_1,\ldots,\hat f_N$ are used as reproductions of the attributes.

The next three sections deal with volumetric B-splines of orders 1, 2, and higher orders.

\subsection{Constant B-Splines}
\label{sec:constant_splines}

In this subsection, we treat the case of volumetric B-splines of order $p=1$, or {\em constant} B-splines.  We show that constant B-splines are equivalent to the Region Adaptive Haar Transform (RAHT) introduced in \cite{QueirozC:16,ChouROQ:16}.  To show their equivalence, we first review RAHT.

\begin{figure}[!t]
\centering
\includegraphics[width=1.0\linewidth]{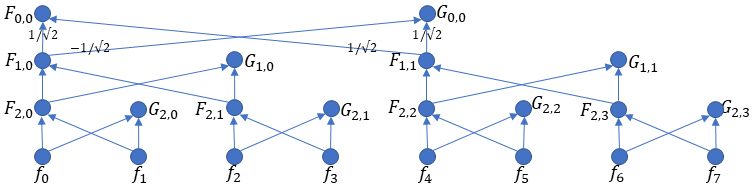}
\caption{Haar Transform butterfly structure for depth $d=3$.}
\label{fig:butterfly}
\end{figure}
\begin{figure}[!t]
\centering
\includegraphics[width=1.0\linewidth]{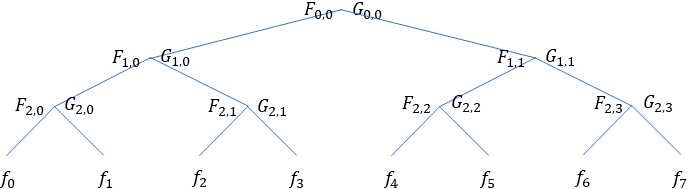}
\caption{Haar Transform tree structure for depth $d=3$.}
\label{fig:HT}
\end{figure}

The Region Adaptive Haar Transform was introduced as a generalization of the Haar Transform.  The Haar Transform of a sequence of $2^d$ coefficients $f_0,\ldots,f_{2^d-1}$ is frequently described as a series of orthonormal butterfly transforms,
\begin{equation}
\left[\begin{array}{c}
\bar F_{\ell,n} \\ \bar G_{\ell,n}
\end{array}\right]
=
\left[\begin{array}{cc}
\frac{1}{\sqrt{2}} & \frac{1}{\sqrt{2}} \\
-\frac{1}{\sqrt{2}} & \frac{1}{\sqrt{2}}
\end{array}\right]
\left[\begin{array}{c}
\bar F_{\ell+1,2n} \\ \bar F_{\ell+1,2n+1}
\end{array}\right],
\end{equation}
for $\ell=d-1,\ldots,0$ and $n=0,\ldots,2^{\ell}-1$, beginning with $\bar F_{d,n}=f_n$, $n=0,\ldots,2^d-1$,  The butterfly structure for $d=3$ is shown in Fig.~\ref{fig:butterfly}.  This can equally well be regarded as a full binary tree, as shown in Fig.~\ref{fig:HT}, in which the signal samples $f_0,\ldots,f_{2^d-1}$ are located at the leaves of the tree, the high pass coefficients $\bar G_{\ell,n}$ are located at the intermediate nodes of the tree, and the DC coefficient is located at the root of the tree.

\begin{figure}[!t]
\centering
\includegraphics[width=1.0\linewidth]{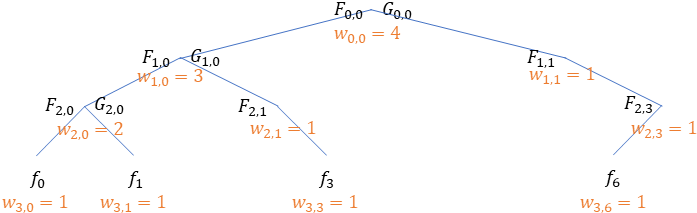}
\caption{RAHT tree structure for depth $d=3$.}
\label{fig:RAHT}
\end{figure}

The Region Adaptive Haar Transform is a generalization of the Haar Transform in that the tree is not necessarily full, in that the internal nodes of the tree may be either binary or unary, and the butterfly transform at each internal binary node of the tree is a Givens rotation,
\begin{equation}
\left[\begin{array}{c}
\bar F_{\ell,n} \\ \bar G_{\ell,n}
\end{array}\right]
=
\left[\begin{array}{cc}
a & b \\
-b & a
\end{array}\right]
\left[\begin{array}{c}
\bar F_{\ell+1,2n} \\ \bar F_{\ell+1,2n+1}
\end{array}\right],
\label{eqn:givens}
\end{equation}
where
\begin{equation}
a = \frac{\sqrt{w_{\ell+1,2n}}}{\sqrt{w_{\ell+1,2n}+w_{\ell+1,2n+1}}},
\end{equation}
\begin{equation}
b = \frac{\sqrt{w_{\ell+1,2n+1}}}{\sqrt{w_{\ell+1,2n}+w_{\ell+1,2n+1}}},
\end{equation}
\begin{equation}
w_{\ell,n} = w_{\ell+1,2n}+w_{\ell+1,2n+1}
\end{equation}
for $\ell=0,\ldots,d-1$, and $w_{d,n}$ equals 1 for all $n$ in the signal and equals 0 otherwise.
$W_{\ell,n}$ is called the {\em weight} of node $n$ at level $\ell$ and is equal to the total of the weights of all the leaves descended from node $n$ in level $\ell$.
The tree for RAHT is shown in Fig.~\ref{fig:RAHT}.

To apply RAHT to a signal $f_1,\ldots,f_N\in\mathbb{R}$ defined on point locations $\bm{x}_1,\ldots,\bm{x}_N\in\mathbb{R}^3$,
first scale the point locations so that the set $\mathcal{X}=\{\bm{x}_1,\ldots,\bm{x}_N\}$ fits within the unit cube, $\mathcal{X}\subset[0,1)^3$, and choose $d$ sufficiently large so that each point location $\bm{x}=(x,y,z)\in\mathcal{X}$ can be represented uniquely with $d$ bits of precision, as $x=\sum_{b=1}^dx_b2^{-b}$, $y=\sum_{b=1}^dy_b2^{-b}$, and $z=\sum_{b=1}^dz_b2^{-b}$, or in more conventional notation,
\begin{eqnarray}
x & = & .x_1\cdots x_d\;, \\
y & = & .y_1\cdots y_d\;, \\
z & = & .z_1\cdots z_d\;.
\end{eqnarray}
The {\em Morton code} of point location $\bm{x}=(x,y,z)$ is defined as the interleaving of its coefficients' bits,
\begin{equation}
\mbox{Morton}(\bm{x}) = .x_1y_1z_1\cdots x_dy_dz_d .
\end{equation}
Now the RAHT tree of depth $3d$ can be constructed with $N$ leaves, with $f_i$ at leaf $i$, such that the path from the root of the tree to leaf $i$ is given by the Morton code of $\bm{x}_i$.  Thus each node in the tree, if it is at level $\ell$, is associated with the common length-$\ell$ prefix shared by the Morton codes of the node's descendants.  Two nodes at level $\ell+1$ are siblings if their length-$\ell$ Morton prefixes are identical.  The Givens rotation is applied to such siblings.


As described above, RAHT is a discrete transform of a signal defined on a discrete set of points.  But it also has an interpretation as a continuous transform of a signal defined on a volume.  To see that, note that the nodes at level $\ell$ of the RAHT tree partition not only the points $\bm{x}_1,\ldots,\bm{x}_N$, but also the volume $[0,1)^3$.  For example, consider a node at a level $\ell=3\ell'$ whose Morton prefix is $.x_1y_1z_1\cdots x_{\ell'}y_{\ell'}z_{\ell'}$.  This node corresponds to the set of all points $\bm{x}\in[0,1)^3$ that have this same Morton prefix, namely the cube
\begin{eqnarray}
\mathcal{B}_{\ell,(n_x,n_y,n_z)}
& = & [2^{-\ell'}n_x,2^{-\ell'}(n_x+1)) \nonumber \\
& \times & [2^{-\ell'}n_y,2^{-\ell'}(n_y+1)) \nonumber \\
& \times & [2^{-\ell'}n_z,2^{-\ell'}(n_z+1)),
\end{eqnarray}
where $n_x=\sum_{b=1}^{\ell'} x_b2^{\ell'-b}$, $n_y=\sum_{b=1}^{\ell'} y_b2^{\ell'-b}$, and $n_z=\sum_{b=1}^{\ell'} z_b2^{\ell'-b}$.  This is a $2^{-\ell'}\times 2^{-\ell'}\times 2^{-\ell'}$ cube at vector integer shift $\bm{n}=(n_x,x_y,n_z)$, where $n_x,n_y,n_z\in\{0,\ldots,2^{\ell'}-1\}$.  For nodes at levels $\ell$ that are not a multiple of three, the notation is a little clumsy, but similar and straightforward.  Specifically, a node at level $\ell=3\ell'+1$ corresponds to a $2^{-(\ell'+1)}\times 2^{-\ell'}\times 2^{-\ell'}$ cuboid $\mathcal{B}_{\ell,\bm{n}}$ at vector integer shift $\bm{n}=(n_x,x_y,n_z)$, where $n_x\in\{0,\ldots,2^{\ell'+1}-1\}$ and $n_y,n_z\in\{0,\ldots,2^{\ell'}-1\}$, and a node at level $\ell=3\ell'+2$ corresponds to a $2^{-(\ell'+1)}\times 2^{-(\ell'+1)}\times 2^{-\ell'}$ cuboid $\mathcal{B}_{\ell,\bm{n}}$ at vector integer shift $\bm{n}=(n_x,x_y,n_z)$, where $n_x,n_y\in\{0,\ldots,2^{\ell'+1}-1\}$ and $n_z\in\{0,\ldots,2^{\ell'}-1\}$.  At any level, we call these cuboids {\em blocks}.  A block $\mathcal{B}_{\ell,\bm{n}}$ is said to be {\em occupied} if it contains a point, that is, if $\mathcal{B}_{\ell,\bm{n}}\cap\mathcal{X}\neq\emptyset$.  Each node in the RAHT tree at level $\ell$ correspond to an occupied block at level $\ell$.

Let $\phi_{\ell,\bm{n}}(\bm{x})$ the the indicator function for block $\mathcal{B}_{\ell,\bm{n}}$, namely
\begin{equation}
\phi_{\ell,\bm{n}}(\bm{x})
=
\left\{\begin{array}{cc}
1&\bm{x}\in\mathcal{B}_{\ell,\bm{n}}\\
0&\mbox{otherwise}\end{array}\right. ,
\label{eqn:phi_ell_n}
\end{equation}
and, as in (\ref{eqn:N_ell}), let $\mathcal{N}_\ell$ be the set of vector integer shifts $\bm{n}$ such that $\phi_{\ell,\bm{n}}(\bm{x}_i)\neq0$ for some $i=1,\ldots,N$, that is, let $\mathcal{N}_\ell$ be
the set of vector integer shifts $\bm{n}$ such that $\mathcal{B}_{\ell,\bm{n}}$ is occupied.  Then, as in (\ref{eqn:F_ell}), let $\mathcal{F}_\ell$ be the subspace of functions that are linear combinations of $\phi_{\ell,\bm{n}}$ for $\bm{n}\in\mathcal{N}_\ell$,
\begin{equation}
\mathcal{F}_\ell=\left\{f_\ell\in\mathcal{F}\left|\exists\{F_{\ell,\bm{n}}\} \mbox{\;s.t.\;} f_\ell=\sum_{\bm{n}\in\mathcal{N}_\ell} F_{\ell,\bm{n}} \phi_{\ell,\bm{n}}\right.\right\}.
\end{equation}
Let $f:[0,1)^3\rightarrow\mathbb{R}$ be any function that agrees with $f_i$ on $\bm{x}_i$, i.e., $f(\bm{x}_i)=f_i$, $i=1,\ldots,N$, and let $f_\ell^*=\sum_{\bm{n}\in\mathcal{N}_\ell} F_{\ell,\bm{n}}^*\phi_{\ell,\bm{n}}$ be the projection of $f$ onto $\mathcal{F}_\ell$.
As in Subsection~\ref{sec:multiresolution}, let $\Phi_\ell^Tf$ denote the $|\mathcal{N}_\ell|\times1$ vector $[\left<\phi_{\ell,\bm{n}_i},f\right>]$, let $\Phi_\ell^T\Phi_\ell$ denote the $|\mathcal{N}_\ell|\times|\mathcal{N}_\ell|$ matrix $[\left<\phi_{\ell,\bm{n}_i},\phi_{\ell,\bm{n}_j}\right>]$, and let $F_\ell^*$ denote the $|\mathcal{N}_\ell|\times1$  vector $[F_{\ell,\bm{n}_j}^*]$.  If $\Phi_\ell^T\Phi_\ell$ is invertible, then
\begin{equation}
F_\ell^* = (\Phi_\ell^T\Phi_\ell)^{-1}\Phi_\ell^Tf.
\end{equation}
From the definition of $\phi_{\ell,\bm{n}}$ (\ref{eqn:phi_ell_n}) and the inner product (\ref{eqn:innerproduct}), it can be seen that $\left<\phi_{\ell,\bm{n}},f\right>=\sum_{\bm{x}_i\in\mathcal{B}_{\ell,\bm{n}}}f(\bm{x}_i)$ and $\left<\phi_{\ell,\bm{n}},\phi_{\ell,\bm{n}'}\right>$ equals $w_{\ell,\bm{n}}$ if $\bm{n}=\bm{n}'$ and equals 0 otherwise, where $w_{\ell,\bm{n}}=\mu(\mathcal{B}_{\ell,\bm{n}})$ is the number of points in the set $\mathcal{B}_{\ell,\bm{n}}$.  Hence $F_{\ell,\bm{n}}^*$ is the average value of the attributes of the points in $\mathcal{B}_{\ell,\bm{n}}$, namely
\begin{equation}
F_{\ell,\bm{n}}^*
= \frac{\left<\phi_{\ell,\bm{n}},f\right>}{w_{\ell,\bm{n}}}
= \frac{1}{w_{\ell,\bm{n}}} \sum_{\bm{x}_i\in\mathcal{B}_{\ell,\bm{n}}}f(\bm{x}_i),
\label{eqn:haar_F_ell_n_star}
\end{equation}
for $\bm{n}\in\mathcal{N}_\ell$.

To express the two-scale equation for $\phi_{\ell,\bm{n}}$ succinctly regardless of whether $\ell$ is a multiple of three or not, we use the following notation. If $\ell,\bm{n}$ are the level and shift of a block, then ``$\ell+1,2\bm{n}$'' and ``$\ell+1,2\bm{n}+1$'' mean the level and shifts of its two subblocks.  To be pedantic, ``$\ell+1,2\bm{n}$'' and ``$\ell+1,2\bm{n}+1$'' mean $\ell+1,(2n_x,n_y,n_z)$ and $\ell+1,(2n_x+1,n_y,n_z)$ if $\ell=3\ell'$ (i.e., $\ell$ is a multiple of 3); they mean $\ell+1,(n_x,2n_y,n_z)$ and $\ell+1,(n_x,2n_y+1,n_z)$ if $\ell=3\ell'+1$ (i.e., $\ell\equiv1\;\mbox{mod}\;3$); and they mean $\ell+1,(n_x,n_y,2n_z)$ and $\ell+1,(n_x,n_y,2n_z+1)$ if $\ell=3\ell'+2$ (i.e., $\ell\equiv2\;\mbox{mod}\;3$).

Now the two-scale equation for $\phi_{\ell,\bm{n}}$ can be easily expressed
\begin{equation}
\phi_{\ell,\bm{n}} = \phi_{\ell+1,2\bm{n}} + \phi_{\ell+1,2\bm{n}+1},
\label{eqn:haar_two_scale}
\end{equation}
so that combining (\ref{eqn:haar_F_ell_n_star}) and (\ref{eqn:haar_two_scale}),
\begin{eqnarray}
F_{\ell,\bm{n}}^*
& = & \frac{\left<\phi_{\ell+1,2\bm{n}},f\right>}{w_{\ell,\bm{n}}} + \frac{\left<\phi_{\ell+1,2\bm{n}+1},f\right>}{w_{\ell,\bm{n}}} \\
& = & \frac{w_{\ell+1,2\bm{n}}}{w_{\ell,\bm{n}}}F_{\ell+1,2\bm{n}}^* + \frac{w_{\ell+1,2\bm{n}+1}}{w_{\ell,\bm{n}}}F_{\ell+1,2\bm{n}+1}^* \\
& = & \frac{w_0}{w_0+w_1}F_{\ell+1,2\bm{n}}^* + \frac{w_1}{w_0+w_1}F_{\ell+1,2\bm{n}+1}^* ,
\label{eqn:haar_F_ell_n_star_two_scale}
\end{eqnarray}
where to be more concise we have abbreviated $w_0=w_{\ell+1,2\bm{n}}$ and $w_1=w_{\ell+1,2\bm{n}+1}$.  Both $w_0$ and $w_1$ will be non-zero when $\ell,\bm{n}$ correspond to a node in the tree with two children.  For such $\ell,\bm{n}$, define the function
\begin{equation}
\psi_{\ell,\bm{n}} = -\frac{\phi_{\ell+1,2\bm{n}}}{w_0} + \frac{\phi_{\ell+1,2\bm{n}+1}}{w_1},
\label{eqn:haar_psi}
\end{equation}
and define
\begin{equation}
G_{\ell,\bm{n}}^* = \frac{w_0w_1}{w_0+w_1}\left<\psi_{\ell,\bm{n}},f\right>.
\label{eqn:haar_G_ell_n_star}
\end{equation}
Then combining (\ref{eqn:haar_G_ell_n_star}), (\ref{eqn:haar_psi}), and (\ref{eqn:haar_F_ell_n_star}),
\begin{align}
G_{\ell,\bm{n}}^*
& = & \frac{w_0w_1}{w_0+w_1}\left(-\frac{\left<\phi_{\ell+1,2\bm{n}},f\right>}{w_0} + \frac{\left<\phi_{\ell+1,2\bm{n}+1},f\right>}{w_1}\right) \\
& = & \frac{w_0w_1}{w_0+w_1}\left(-F_{\ell+1,2\bm{n}}^* + F_{\ell+1,2\bm{n}+1}^*\right),
\label{eqn:haar_G_ell_n_star_two_scale}
\end{align}
which is the scaled difference between the average values of the attributes of the points in the two sub-blocks $\mathcal{B}_{\ell+1,2\bm{n}}$ and $\mathcal{B}_{\ell+1,2\bm{n}+1}$ of $\mathcal{B}_{\ell,\bm{n}}$.  When $f$ is smooth, $G_{\ell,\bm{n}}^*$ will be close to zero.  Putting (\ref{eqn:haar_F_ell_n_star_two_scale}) and (\ref{eqn:haar_G_ell_n_star_two_scale}) in matrix form,
\begin{equation}
\left[\begin{array}{c}
F_{\ell,\bm{n}}^*\\G_{\ell,\bm{n}}^*
\end{array}\right]
=
\left[\begin{array}{cc}
\frac{w_0}{w_0+w_1} & \frac{w_1}{w_0+w_1} \\
-\frac{w_0w_1}{w_0+w_1} & \frac{w_0w_1}{w_0+w_1}
\end{array}\right]
\left[\begin{array}{c}
F_{\ell+1,2\bm{n}}^*\\F_{\ell+1,2\bm{n}+1}^*
\end{array}\right].
\label{eqn:haar_nonnormalized}
\end{equation}

Clearly both $\phi_{\ell,\bm{n}}$ and $\psi_{\ell,\bm{n}}$ have support only on $\mathcal{B}_{\ell,\bm{n}}$, and hence they are both orthogonal to both $\phi_{\ell,\bm{n}'}$ and $\psi_{\ell,\bm{n}'}$ for vector integer shifts $\bm{n}'\neq\bm{n}$.  But it can also be seen that $\phi_{\ell,\bm{n}}$ and $\psi_{\ell,\bm{n}}$ are orthogonal to each other, since $\left<\phi_{\ell,\bm{n}},\psi_{\ell,\bm{n}}\right>=-1+1=0$.  Thus the orthogonal complement of $\mathcal{F}_\ell$ in $\mathcal{F}_{\ell+1}$ is
\begin{equation}
\mathcal{G}_\ell
=
\left\{
g_\ell\left|
g_\ell = \sum_{\bm{n}\in\mathcal{N}_\ell^b}
G_{\ell,\bm{n}}\psi_{\ell,\bm{n}}
\right.
\right\} ,
\end{equation}
where the sum is over only those vector integer shifts $\mathcal{N}_\ell^b\subseteq\mathcal{N}_\ell$ for which $\ell,\bm{n}$ correspond to a node in the tree with two children.
However, as defined, the orthogonal basis functions $\phi_{\ell,\bm{n}}$, and $\psi_{\ell,\bm{n}}$ are not normalized.  Define their normalized versions
\begin{equation}
\bar\phi_{\ell,\bm{n}} = \frac{\phi_{\ell,\bm{n}}}{||\phi_{\ell,\bm{n}}||} = \frac{\phi_{\ell,\bm{n}}}{\sqrt{w_0+w_1}} ,
\end{equation}
\begin{equation}
\bar\psi_{\ell,\bm{n}} = \frac{\psi_{\ell,\bm{n}}}{||\psi_{\ell,\bm{n}}||} = \frac{\psi_{\ell,\bm{n}}}{\sqrt{\frac{1}{w_0}+\frac{1}{w_1}}}
= \frac{\sqrt{w_0w_1}}{\sqrt{w_0+w_1}}\psi_{\ell,\bm{n}} .
\end{equation}
Then $f_\ell=\sum_{\bm{n}\in\mathcal{N}_\ell}F_{\ell,\bm{n}}\phi_{\ell,\bm{n}}=\sum_{\bm{n}\in\mathcal{N}_\ell}\bar F_{\ell,\bm{n}}\bar\phi_{\ell,\bm{n}}$ and  $g_\ell=\sum_{\bm{n}\in\mathcal{N}_\ell^b}G_{\ell,\bm{n}}\psi_{\ell,\bm{n}}=\sum_{\bm{n}\in\mathcal{N}_\ell^b}\bar G_{\ell,\bm{n}}\bar\psi_{\ell,\bm{n}}$, where
\begin{equation}
\bar F_{\ell,\bm{n}} = \sqrt{w_0+w_1} F_{\ell,\bm{n}} ,
\end{equation}
\begin{equation}
\bar G_{\ell,\bm{n}} = \frac{\sqrt{w_0+w_1}}{\sqrt{w_0w_1}} G_{\ell,\bm{n}} .
\end{equation}
Rewriting (\ref{eqn:haar_nonnormalized}),
\begin{equation}
\left[\begin{array}{c}
\frac{1}{\sqrt{w_0+w_1}}\bar F_{\ell,\bm{n}}^* \\ \frac{\sqrt{w_0w_1}}{\sqrt{w_0+w_1}}\bar G_{\ell,\bm{n}}^*
\end{array}\right]
=
\left[\begin{array}{cc}
\frac{w_0}{w_0+w_1} & \frac{w_1}{w_0+w_1} \\
-\frac{w_0w_1}{w_0+w_1} & \frac{w_0w_1}{w_0+w_1}
\end{array}\right]
\left[\begin{array}{c}
\frac{\bar F_{\ell+1,2\bm{n}}^*}{\sqrt{w_0}} \\ \frac{\bar F_{\ell+1,2\bm{n}+1}^*}{\sqrt{w_1}}
\end{array}\right],
\end{equation}
or
\begin{equation}
\left[\begin{array}{c}
\bar F_{\ell,\bm{n}}^* \\ \bar G_{\ell,\bm{n}}^*
\end{array}\right]
=
\left[\begin{array}{cc}
\frac{\sqrt{w_0}}{\sqrt{w_0+w_1}} & \frac{\sqrt{w_1}}{\sqrt{w_0+w_1}} \\
-\frac{\sqrt{w_1}}{\sqrt{w_0+w_1}} & \frac{\sqrt{w_0}}{\sqrt{w_0+w_1}}
\end{array}\right]
\left[\begin{array}{c}
\bar F_{\ell+1,2\bm{n}}^* \\ \bar F_{\ell+1,2\bm{n}+1}^*
\end{array}\right].
\end{equation}
This is identical to (\ref{eqn:givens}) with
\begin{equation}
a = \frac{\sqrt{w_0}}{\sqrt{w_0+w_1}} = \frac{\sqrt{w_{\ell+1,2\bm{n}}}}{\sqrt{w_{\ell+1,2\bm{n}}+w_{\ell+1,2\bm{n}+1}}},
\end{equation}
\begin{equation}
b = \frac{\sqrt{w_1}}{\sqrt{w_0+w_1}}  = \frac{\sqrt{w_{\ell+1,2\bm{n}+1}}}{\sqrt{w_{\ell+1,2\bm{n}}+w_{\ell+1,2\bm{n}+1}}}.
\end{equation}
The asterisks remind us that if we apply RAHT to the values $f_1,\ldots,f_N$ of a volumetric function $f(\bm{x})$ at point locations $\bm{x}_1,\ldots,\bm{x}_N\in\mathbb{R}^3$, the resulting coefficients $\bar F_{\ell,\bm{n}}^*$, $\bm{n}\in\mathcal{N}_\ell$, are optimal in that they represent the projection $f^*=\sum_{\bm{n}\in\mathcal{N}_\ell}\bar F_{\ell,\bm{n}}^*\phi_{\ell,\bm{n}}$ of $f$ onto the subspace $\mathcal{F}_\ell$.

Thus RAHT has a volumetric interpretation.



\subsection{Tri-linear B-Splines}
\label{sec:linear_splines}

In this subsection, we treat the case of volumetric B-splines of order $p=2$, or {\em tri-linear} B-splines.  These splines are continuous, unlike
constant B-splines.  Tri-linear B-splines reduce blocking artifacts, and perhaps more importantly, do not develop rips, tears, or holes when the surface or motion representation is quantized, due to their guaranteed continuity.

We begin with the central cardinal B-spline of order $p=2$,
\begin{equation}
\phi^{(2)}(x-1)=\left\{\begin{array}{cc}
1+x & \mbox{$x\in[-1,0]$} \\
1-x & \mbox{$x\in[0,1]$} \\
0 & \mbox{otherwise}
\end{array}\right.,
\label{eqn:hat1}
\end{equation}
and define the volumetric version
\begin{equation}
\phi_{0,\bm{0}}(\bm{x})=\phi^{(2)}(x-1)\phi^{(2)}(y-1)\phi^{(2)}(z-1).
\label{eqn:hat2}
\end{equation}
Then
\begin{equation}
\phi_{\ell,\bm{n}}(\bm{x})=\phi_{0,\bm{0}}(2^\ell\bm{x}-\bm{n})
\label{eqn:hat3}
\end{equation}
is the tri-linear B-spline basis function at level $\ell$ with vector integer shift $\bm{n}$.  As in (\ref{eqn:F_ell}), let
\begin{equation}
\mathcal{F}_\ell=\left\{f_\ell\in\mathcal{F}\left|\exists\{F_{\ell,\bm{n}}\} \mbox{\;s.t.\;} f_\ell(\bm{x})=\sum_{\bm{n}\in\mathcal{C}_\ell} F_{\ell,\bm{n}} \phi_{\ell,\bm{n}}(\bm{x})\right.\right\}.
\label{eqn:F_ell3}
\end{equation}
be the subspace of all functions in the Hilbert space $\mathcal{F}$ that are tri-linear over all blocks $\mathcal{B}_{\ell,\bm{n}}$, $\bm{n}\in\mathbb{Z}^3$, at level $\ell$.  It suffices to take the sum over vector integer shifts $\bm{n}\in\mathcal{C}_\ell$, where $\mathcal{C}_\ell$ is the collection of {\em corners} of the occupied blocks $\mathcal{B}_{\ell,\bm{n}'}$, $\bm{n}'\in\mathcal{N}_\ell$.  This is because $\phi_{\ell,\bm{n}}(\bm{x}_i)=0$ for all vector integer shifts $\bm{n}\not\in\mathcal{C}_\ell$.

As in (\ref{eqn:F_ell_star}), the projection of $f\in\mathcal{F}$ onto $\mathcal{F}_\ell$ is given by $F_\ell^* = (\Phi_\ell^T\Phi_\ell)^{-1}\Phi_\ell^Tf$, where $\Phi_\ell^Tf$ is the $|\mathcal{C}_\ell|\times1$ vector $[\left<\phi_{\ell,\bm{n}_i},f\right>]$, $\Phi_\ell^T\Phi_\ell$ is the $|\mathcal{C}_\ell|\times|\mathcal{C}_\ell|$ matrix $[\left<\phi_{\ell,\bm{n}_i},\phi_{\ell,\bm{n}_j}\right>]$, and $F_\ell^*$ is the $|\mathcal{C}_\ell|\times1$  vector $[F_{\ell,\bm{n}_j}^*]$.  However, the matrix $\Phi_\ell^T\Phi_\ell$ in the case of tri-linear splines, unlike the case of constant splines, is block tri-diagonal rather than diagonal, and hence is not so trivial to invert.  Nevertheless, $\Phi_\ell^T\Phi_\ell$ is sparse, and thus inversion by iterative methods is quite feasible.

The elements of $\Phi_\ell^T\Phi_\ell$ and $\Phi_\ell^Tf$ do not have to be computed directly from their definitions.  Rather, they can be computed directly from their definitions in the simple case of $\ell=d$, and then they can be computed recursively for $\ell<d$, using the two-scale equation (\ref{eqn:twoscale}).  For the tri-linear B-spline, the coefficients in the two-scale equation are
\begin{equation}
a_{\bm{k}}
=\left\{\begin{array}{cc}
2^{-||\bm{k}||_1} & \bm{k}\in\{-1,0,1\}^3 \\
0 & \mbox{otherwise}
\end{array}\right. ,
\end{equation}
where $||\bm{k}||_1=|k_x|+|k_y|+|k_z|$ is the 1-norm of $\bm{k}=(k_x,k_y,k_z)$. Thus, as in (\ref{eqn:phi_ell_T_f}) and (\ref{eqn:phi_ell_T_phi_ell}), for $\ell<d$,
\begin{equation}
\Phi_\ell^Tf = A_\ell\Phi_{\ell+1}^Tf
\end{equation}
and
\begin{equation}
\Phi_\ell^T\Phi_\ell = A_\ell\Phi_{\ell+1}^T\Phi_{\ell+1}A_\ell^T,
\end{equation}
where $A_\ell=[a_{\bm{n}_j-2\bm{n}_i}]$ for $\bm{n}_i\in\mathcal{C}_\ell$, $\bm{n}_j\in\mathcal{C}_{\ell+1}$.  For $\ell=d$, we may use simply
\begin{equation}
\Phi_d^Tf = [f_i]
\end{equation}
and
\begin{equation}
\Phi_d^T\Phi_d = I_N,
\end{equation}
where $I_N$ is the $N\times N$ identity matrix.  This is possible because the point locations $\bm{x}_i$ can be taken to be the origins of the occupied {\em voxels}, or blocks $\mathcal{B}_{d,\bm{n}}$ at level $d$.

Consider now the orthogonal complement $\mathcal{G}_\ell$ of $\mathcal{F}_\ell$ in $\mathcal{F}_{\ell+1}$.  As in the case of constant B-splines, in the case of tri-linear B-splines, the basis functions $\psi_{\ell,\bm{n}}$ of $\mathcal{G}_\ell$ depend locally on the point locations $\{\bm{x}_i\}$, hence are not shifts of each other as in the case of Lebesgue measure.  In the case of constant B-splines, we were able to define explicitly basis functions $\psi_{\ell,\bm{n}}$ orthogonal to each other and to $\mathcal{F}_\ell$.  Unfortunately, in the case of tri-linear B-splines, this is less easy to do.
Nevertheless, for tri-linear B-splines, the procedure in Section~\ref{sec:wavelets} may be followed for constructing a basis for $\mathcal{G}_\ell$.  For smoothing applications in which high pass coefficients are simply set to zero, this procedure is sufficient.  However, for coding applications, it is additionally important to orthonormalize the basis functions, so that scalar quantization of their coefficients does not introduce more error in the signal domain than in the transform domain.  To orthonormalize the basis functions, the procedure in Section~\ref{sec:orthgonality} may be followed.



\subsection{Higher-Order B-Splines}
\label{sec:higher_order_splines}

Volumetric B-splines of order $p\geq 3$ are possible, and offer higher order continuity properties.  However, they are more complex to compute, and at each level $\ell$, they require significantly more coefficients per occupied block.  Specifically, a function in $\mathcal{F}_\ell$ requires $p^3$ coefficients per occupied block.  Although coefficients beyond the first $N$ need not be transmitted as they are linear combinations of the first $N$, extra smoothness gained from the higher order may not be worth the added complexity.

\section{Geometry Coding using B\'ezier Volumes}
\label{sec:geometry}

In the last section, we showed how to represent and compress attributes $f_1,\ldots,f_N$ on a given set of point locations $\bm{n}_1,\ldots,\bm{n}_N$.  The point locations were assumed to be communicated out-of-band as side information and reconstructed at the decoder, either with or without loss.

In this section, we show how to represent and compress the point locations themselves.  Our approach is to represent the point locations, or more generally the point cloud {\em geometry}, implicitly as the level set of a volumetric function.  The volumetric function, now a proxy for the geometry, can be represented as a tri-linear B-spline, decomposed into a low-resolution ``approximation'' and a set of high-pass coefficients (wavelet coefficients), and coded.

\subsection{Implicit Surfaces}

Let $f:\mathbb{R}^3\rightarrow\mathbb{R}$ be a volumetric function and let $c$ be a constant.  The set of points $\bm{x}$ satisfying the equation $f(\bm{x})=c$ is called the {\em level set} of $f$ at level $c$, or the $c$-level set of $f$.  The $c$-level set $\mathcal{S}=\{\bm{x}\left|f(\bm{x})=c\right.\}$ of $f$ defines
a set in $\mathbb{R}^3$ called an {\em implicit surface}.

Let $\Omega$ be a set in $\mathbb{R}^3$, and let $\mathcal{S}=\partial\Omega$ be its boundary.\footnote{The {\em boundary} of a set in $\mathbb{R}^3$ is the set of points that are limit points both of the set and of its complement.}  Informally, we will refer to the boundary of a set as its {\em surface}.  The 
{\em signed distance function} (SDF) of $\Omega$ is the function\footnote{Note that in \eqref{eqn:SDF} we use the SDF sign convention where a positive sign indicates that the point is outside the set and a negative sign indicates that the point is inside the set, following \cite{Osher2003}.}
\begin{equation}
f(\bm{x})=\left\{\begin{array}{cc}
-d(\bm{x},\mathcal{S}) & \mbox{if $\bm{x}\in\Omega$} \\
d(\bm{x},\mathcal{S}) & \mbox{otherwise} ,
\end{array}\right.
\label{eqn:SDF}
\end{equation}
where $d(\bm{x},\mathcal{S})=\min_{\bm{x}_0\in\mathcal{S}}||\bm{x}-\bm{x}_0||$ is the distance from $\bm{x}$ to its closest point in the set $\mathcal{S}$.\footnote{Since $\mathcal{S}$ is closed, the minimum is well-defined and is achieved by a point $\bm{x}_0^*\in\mathcal{S}$.}  Clearly, $f(\bm{x})=0$ if and only if $\bm{x}\in\mathcal{S}$.  Thus the boundary of $\Omega$ is given by the 0-level set of $f$,
\begin{equation}
\mathcal{S} = \{\bm{x}|f(\bm{x})=0\}.
\end{equation}

If the function $f$ can be approximated by a function $\hat f$, then the surface $\mathcal{S}$ can be approximated by the 0-level set of $\hat f$,
\begin{equation}
\widehat{\mathcal{S}} = \{\bm{x}|\hat f(\bm{x})=0\}.
\label{eqn:S_hat}
\end{equation}

Commonly, the surface $\mathcal{S}$ is locally planar.  This means that near a point $\bm{x}_0$ on the surface, the signed distance function is approximated by $(\bm{x}-\bm{x}_0)\cdot\bm{n}_0$, where $\bm{n}_0$ is the unit normal to the surface at $\bm{x}_0$.  Thus, near the surface, the signed distance function is nearly tri-linear and hence is well-approximated by a function in the subspace of tri-linear B-splines at a sufficiently high level of detail.

Our strategy is to approximate $f$ by functions $f_\ell$, $\ell=0,1,\ldots$ in a nested sequence of subspaces $\mathcal{F}_\ell$ (\ref{eqn:nested}) of tri-linear B-splines (\ref{eqn:hat1})--(\ref{eqn:hat3}).  At a sufficiently high level of detail, say $\ell$, the energy $||f-f_\ell||^2$ is low, meaning that the approximation $f_\ell$ is very close to $f$, as the surface is approximately planar over patches of diameter $2^{-\ell}$ or smaller. The parameters of $f_\ell$ can then be quantized, entropy coded, and transmitted.  The reconstructed function $\hat f=\hat f_\ell$, despite its quantized parameters, remains in $\mathcal{F}_\ell$, and hence is guaranteed to be continuous.  Importantly, the reconstructed surface $\hat{\mathcal{S}}$ (\ref{eqn:S_hat}) is continuous in the sense that no holes develop as an artifact of quantization.

Computation of the function $f(\bm{x})$ is often an integral part of the processing pipeline used to produce the point cloud.  For example, Curless and Levoy \cite{CurlessL:96} compute a (truncated) signed distance function $f(\bm{x})$ of an object (from a set of depth camera measurements $\mathcal{M}$) in order to reconstruct the surface of the object as the 0-level set of $f$.  Similarly, Loop et al.\ \cite{LoopCOC:16} compute the occupancy probability $p(\bm{x})=P\{\bm{x}\in\Omega|\mathcal{M}\}$ in order to reconstruct the surface of an object as the (0.5)-level set of $p$, or equivalently, the $0$-level set of the log odds $l(\bm{x})=\log p(\bm{x})/(1-p(\bm{x}))$.  (The log odds $l(\bm{x})$ has a better linear approximation than $p(\bm{x})$ in the vicinity of the surface, and thus is easier to approximate as a tri-linear B-spline.)

\begin{figure}[!h]
\centering
\includegraphics[width=1\linewidth]{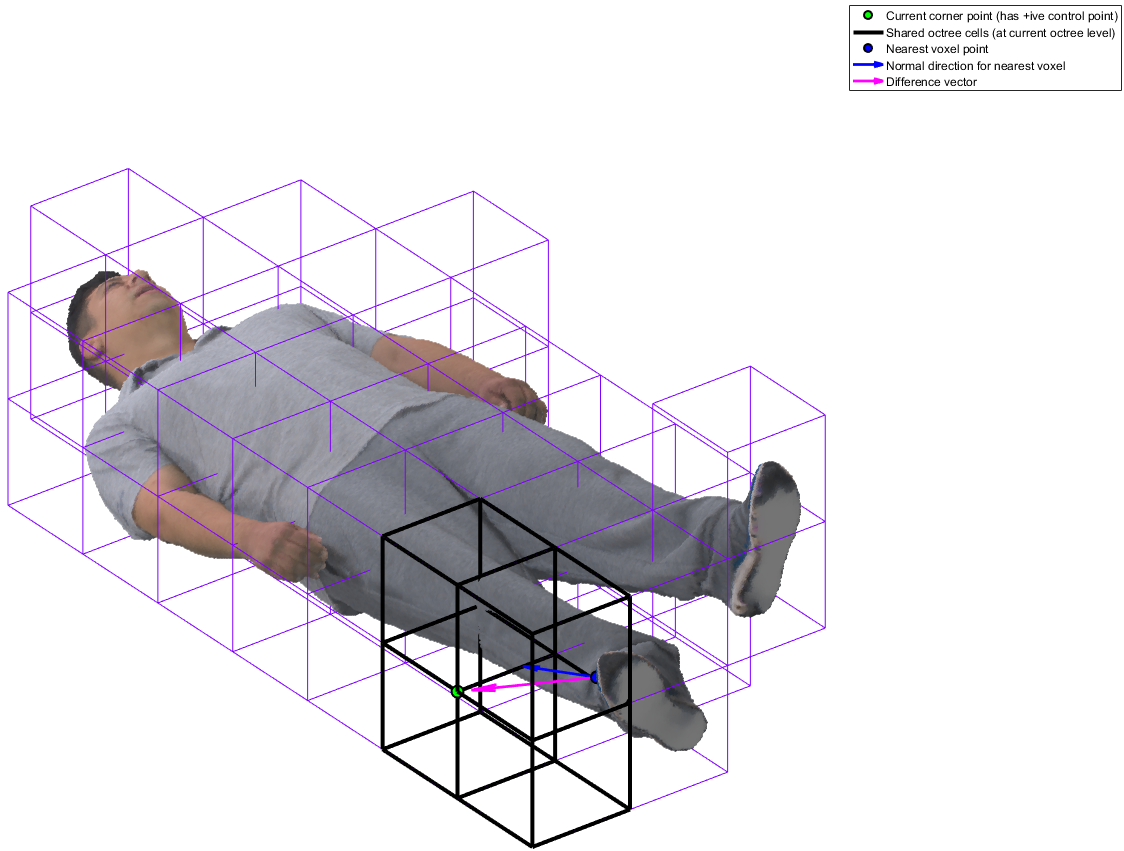}
\caption{Example of a positive SDF value (control point) on a shared octree block corner at level 3.  The example corner, colored in green, is $outside$ the point cloud surface, hence it has a positive SDF value.}
\label{fig:PosSDFExample}
\end{figure}

\begin{figure}[!h]
\centering
\includegraphics[width=1\linewidth]{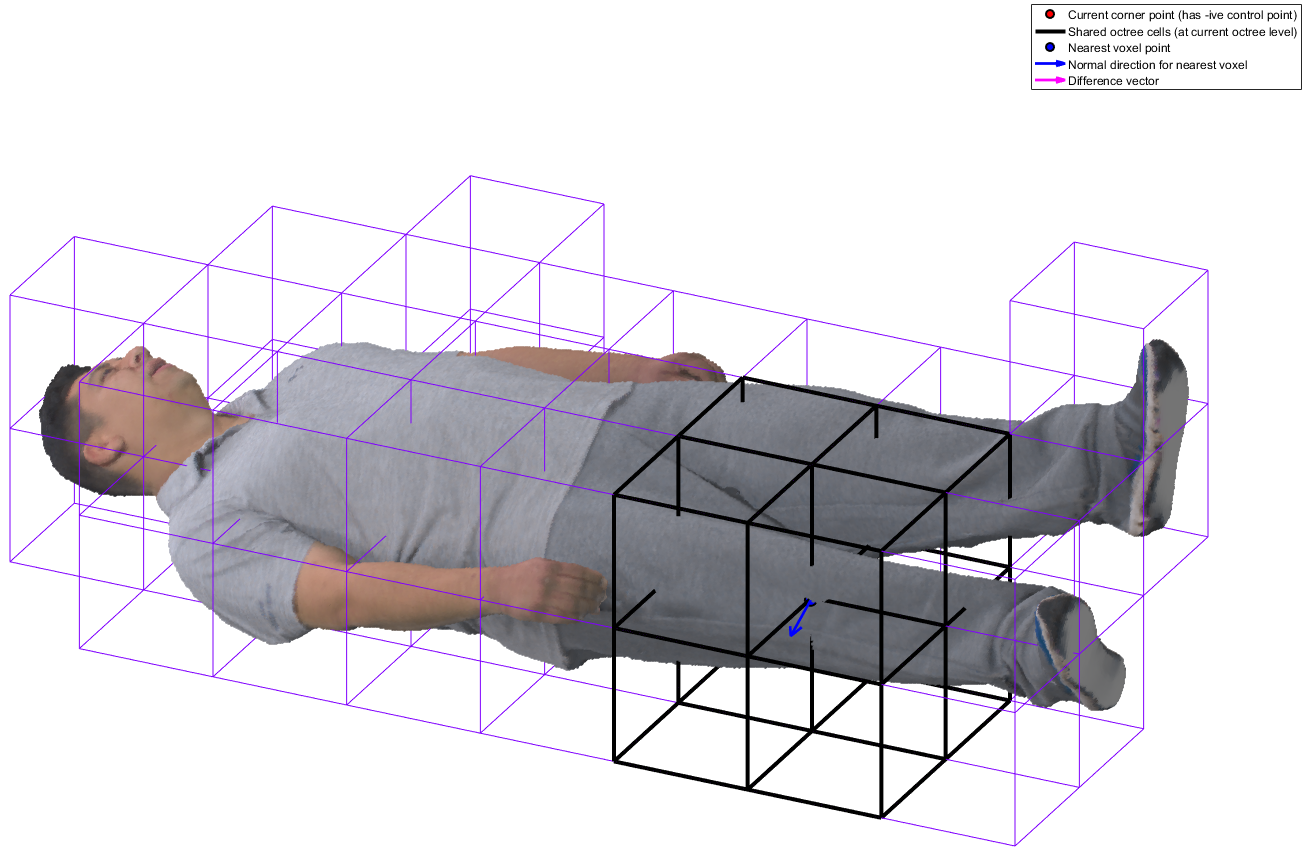}
\includegraphics[width=0.5\linewidth]
{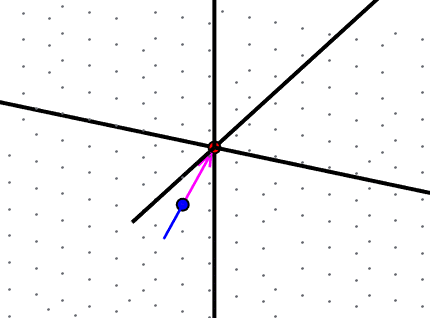}
\caption{Example of a negative SDF value (control point) on a shared octree block corner at level 3.  The bottom image shows a zoomed-in version of the shared corner and its nearest voxel indicated in the top picture, to illustrate that the difference vector between the corner and its nearest voxel is pointing in the opposite direction to the normal vector at that voxel, hence the control point for that corner is negative, as the corner is $inside$ the point cloud surface.}
\label{fig:NegSDFExample}
\end{figure}

In many cases, however, the function $f$ may have to be derived from a finite collection of points $\bm{x}_1,\ldots,\bm{x}_N$ sampled from a surface.  In this case, the signed distance function can be approximated as
\begin{equation}
f(\bm{x})=sign\left((\bm{x}-\bm{x}_{i^*})\cdot\bm{n}_{i^*} \right)\left(\min_i||\bm{x}-\bm{x}_i|| \right),
\label{eqn:sdf1}
\end{equation}
where 
\begin{equation}
i^* = \arg\min_i ||\bm{x}-\bm{x}_i||,
\end{equation}
when $\bm{x}$ is far from the surface, and as
\begin{equation}
f(\bm{x})=(\bm{x}-\bm{x}_{i})\cdot\bm{n}_{i}
\label{eqn:sdf2}
\end{equation}
when $\bm{x}$ is close to a point $\bm{x}_i$ on the surface, with unit normal $\bm{n}_i$. Here we assume that a unit normal at each point can be computed.  Our method for computing the SDF values (or {\em control points}) follows \eqref{eqn:sdf1}-\eqref{eqn:sdf2}. The SDF computation is carried out for all the {\em unique} octree block corners at each octree level, from the root (level 0) to the voxel level (where the block size is $1\times1\times1$).  When checking for the nearest voxel to a corner, we consider only the voxels in the octree blocks (at the same octree level) that share that corner.  We define a positive SDF value to indicate that the corresponding corner is {\em outside} the point cloud ``surface'', while a negative SDF value indicates that the corresponding corner is {\em inside} the surface.  As indicated in \eqref{eqn:sdf1}, a control point will be positive if the normal vector of the nearest voxel to that corner, and the difference vector between the corner and that nearest voxel, are pointing in a similar direction (i.e., the normal vector points {\em towards} the corner), while a control point will be negative if the normal vector and the difference vector point in different directions (i.e., the normal vector points {\em away} from the corner).  Fig.~\ref{fig:PosSDFExample} and Fig.~\ref{fig:NegSDFExample} show examples of a positive and negative SDF value (control point), respectively, at octree level 3.  Note that the {\em difference vector} in Fig.~\ref{fig:PosSDFExample} and Fig.~\ref{fig:NegSDFExample} represents the vector $(\bm{x}-\bm{x}_{i^*})$ from \eqref{eqn:sdf1}, while the {\em normal vector} is the vector $\bm{n}_{i^*}$.

\subsection{Nested B-Spline Spaces}

The nested subspaces of tri-linear B-splines
\begin{equation}
\mathcal{F}_0\subseteq\mathcal{F}_1\subseteq\ldots\subseteq\mathcal{F}_\ell\subseteq\mathcal{F}_{\ell+1}\subseteq\ldots\subseteq\mathcal{F}
\end{equation}
are defined the same way as in (\ref{eqn:F_ell3}), namely
\begin{equation}
\mathcal{F}_\ell=\left\{f_\ell\in\mathcal{F}\left|\exists\{F_{\ell,\bm{n}}\} \mbox{\;s.t.\;} f_\ell(\bm{x})=\sum_{\bm{n}\in\mathcal{C}_\ell} F_{\ell,\bm{n}} \phi_{\ell,\bm{n}}(\bm{x})\right.\right\},
\label{eqn:F_ell2}
\end{equation}
except that in (\ref{eqn:F_ell2}) the space $\mathcal{F}$ is the Hilbert space with the inner product defined with the more usual Lebesgue measure rather than with the counting measure (\ref{eqn:measure}) on the points $\bm{x}_1,\ldots,\bm{x}_N$.  This is because we now care about the value of $f$ not just on the points $\bm{x}_1,\ldots,\bm{x}_N$ but also on points $\bm{x}$ away from the surface, so that $f_\ell(\bm{x})=c$ can be used to approximate an implicit surface.

Since $\phi_{\ell,\bm{n}}(\bm{x})$ is the tri-linear ``hat'' function (\ref{eqn:hat1})--(\ref{eqn:hat3}), for which $\phi_{\ell,\bm{n}}(2^{-\ell}\bm{n})=1$ and $\phi_{\ell,\bm{n}}(2^{-\ell}\bm{k})=0$ for all $\bm{k}\neq\bm{n}$, $\bm{k}\in\mathbb{Z}^3$, it follows that $F_{\ell,\bm{n}}=f_\ell(2^{-\ell}\bm{n})$ for all $\bm{n}\in\mathbb{Z}^3$.  Hence the coefficients representing $f_\ell$ in the basis $\{\phi_{\ell,\bm{n}}\}$ are simply the values of $f_\ell(\bm{x})$ at the corners $\mathcal{C}_\ell$ of the occupied blocks $\mathcal{B}_\ell$ at level $\ell$.

At this point, we could develop the usual decomposition of $\mathcal{F}_{\ell+1}$ into $\mathcal{F}_\ell$ and its orthogonal complement $\mathcal{G}_\ell$, where $\mathcal{G}_\ell$ is spanned by wavelet basis functions that are orthogonal not only to $\mathcal{F}_\ell$ but to each other.  However, these wavelet basis functions would have infinite support, and would be impractical to implement in $\mathbb{R}^3$.

We propose, instead, a non-orthogonal decomposition $\mathcal{F}_{\ell+1}=\mathcal{F}_\ell\oplus\mathcal{G}_\ell$, where
\begin{equation}
\mathcal{G}_\ell=\left\{g_\ell\in\mathcal{F}\left|\exists\{G_{\ell,\bm{n}}\} \mbox{\;s.t.\;} g_\ell=\sum_{\bm{n}\in\mathcal{C}_{\ell+1}\setminus2\mathcal{C}_\ell} G_{\ell,\bm{n}} \phi_{\ell+1,\bm{n}}\right.\right\}.
\label{eqn:G_ell}
\end{equation}
Clearly, $\mathcal{G}_\ell$ is a subspace of $\mathcal{F}_{\ell+1}$ spanned by the functions $\phi_{\ell+1,\bm{n}}$ centered at the corners of the blocks at level $\ell+1$ that are not also corners of any blocks at level $\ell$, namely $\{\phi_{\ell+1,\bm{n}}:\bm{n}\in\mathcal{C}_{\ell+1}\setminus2\mathcal{C}_\ell\}$. Note that just as $F_{\ell,\bm{n}}=f_\ell(2^{-\ell}\bm{n})$ in (\ref{eqn:F_ell2}), $G_{\ell,\bm{n}}=g_\ell(2^{-(\ell+1)}\bm{n})$ in (\ref{eqn:G_ell}), for all $\bm{n}\in\mathbb{Z}^3$.  Thus $\mathcal{G}_\ell$ is the subspace of functions in $\mathcal{F}_{\ell+1}$ that are zero on the corners of the blocks in $\mathcal{B}_\ell$.  
It can be seen that if $f_\ell\in\mathcal{F}_\ell$ and $g_\ell\in\mathcal{G}_\ell$ then $f_\ell+g_\ell$ is a function $f_{\ell+1}\in\mathcal{F}_{\ell+1}$ such that for all $\bm{n}\in2\mathcal{C}_\ell$, $f_{\ell+1}(2^{-(\ell+1)}\bm{n})=f_\ell(2^{-\ell}\bm{n}/2)=F_{\ell,\bm{n}/2}=F_{\ell+1,\bm{n}}$, and for all $\bm{n}\in\mathcal{C}_{\ell+1}\setminus2\mathcal{C}_\ell$, $f_{\ell+1}(2^{-(\ell+1)}\bm{n})=f_\ell(2^{-(\ell+1)}\bm{n})+g_\ell(2^{-(\ell+1)}\bm{n})=f_\ell(2^{-(\ell+1)}\bm{n})+G_{\ell,\bm{n}}=F_{\ell+1,\bm{n}}$.  Thus for all $\bm{n}\in\mathcal{C}_{\ell+1}\setminus2\mathcal{C}_\ell$, $f_\ell(2^{-(\ell+1)}\bm{n})$ can be considered the {\em prediction} of $f_{\ell+1}(2^{-(\ell+1)}\bm{n})$, and $G_{\ell,\bm{n}}$ can be considered the {\em prediction error} or {\em residual}.
Thus $\mathcal{F}_{\ell+1}$ is the direct sum of $\mathcal{F}_\ell$ and $\mathcal{G}_\ell$.

This implies
\begin{equation}
\mathcal{F}_{\ell+1} = \mathcal{F}_0\oplus\mathcal{G}_0\oplus\mathcal{G}_1\oplus\cdots\oplus\mathcal{G}_\ell,
\end{equation}
and hence any function $f_{\ell+1}\in\mathcal{F}_{\ell+1}$ can be written
\begin{equation}
f_{\ell+1} = f_0 + g_0 + g_1 + \cdots + g_\ell,
\end{equation}
where the coefficients $\{F_{0,\bm{n}}\}$ of $f_0$ in the basis for $\mathcal{F}_0$ are {\em low pass} coefficients, while the coefficients $\{G_{\ell,\bm{n}}\}$ of $g_\ell$ in the basis for $\mathcal{G}_\ell$ are {\em high pass} or {\em wavelet} coefficients.

Any function $f:\mathbb{R}^3\rightarrow\mathbb{R}$ can be projected onto $\mathcal{F}_\ell$ by sampling:
\begin{equation}
F_{\ell,\bm{n}}=f_\ell(2^{-\ell}\bm{n})=f(2^{-\ell}\bm{n})
\label{eqn:sampling_projection}
\end{equation}
for all $\bm{n}\in\mathbb{Z}^3$.  That is, the projection of $f$ onto $\mathcal{F}_\ell$, denoted $f_\ell=f\circ\mathcal{F}_\ell$, interpolates $f$ at the knots $2^{-\ell}\mathbb{Z}^3$.  Note that this definition of projection is not the least squares projection of $f$ onto $\mathcal{F}_\ell$ under the inner product given by Lebesque measure.  However, under this definition of projection, the functions $g_\ell\in\mathcal{G}_\ell$ are orthogonal to $\mathcal{F}_\ell$, in the sense that they are zero at the knots $2^{\-\ell}\mathbb{Z}^3$ and thus project to the zero function in $\mathcal{F}_\ell$.

\begin{figure}[!h]
\centering
\includegraphics[width=0.6\linewidth]{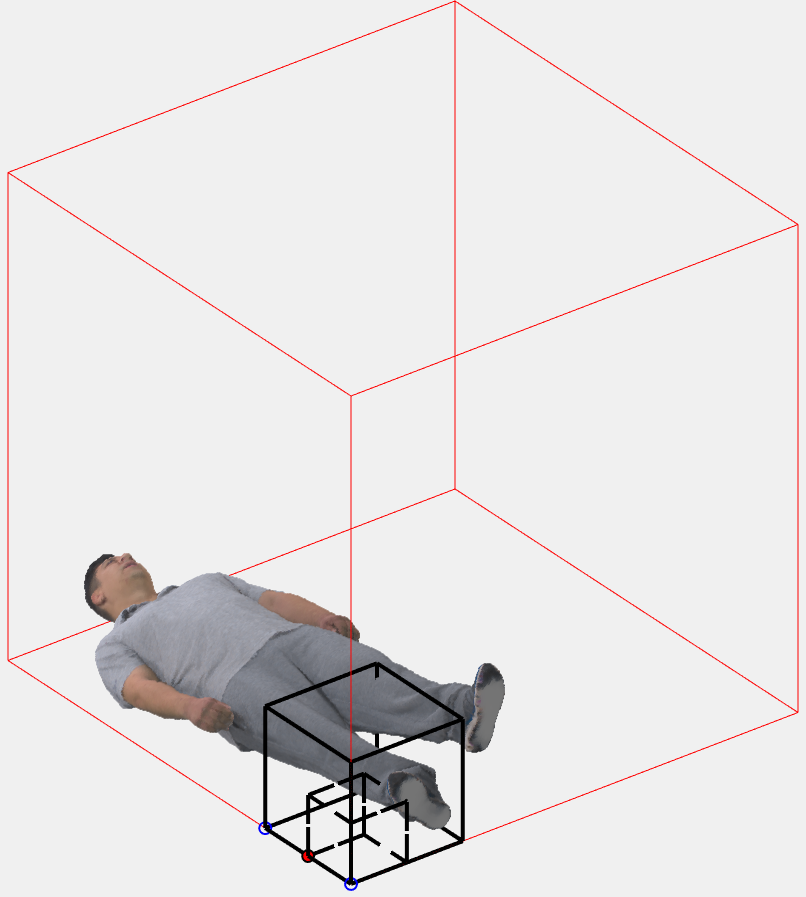}
\includegraphics[width=0.6\linewidth]{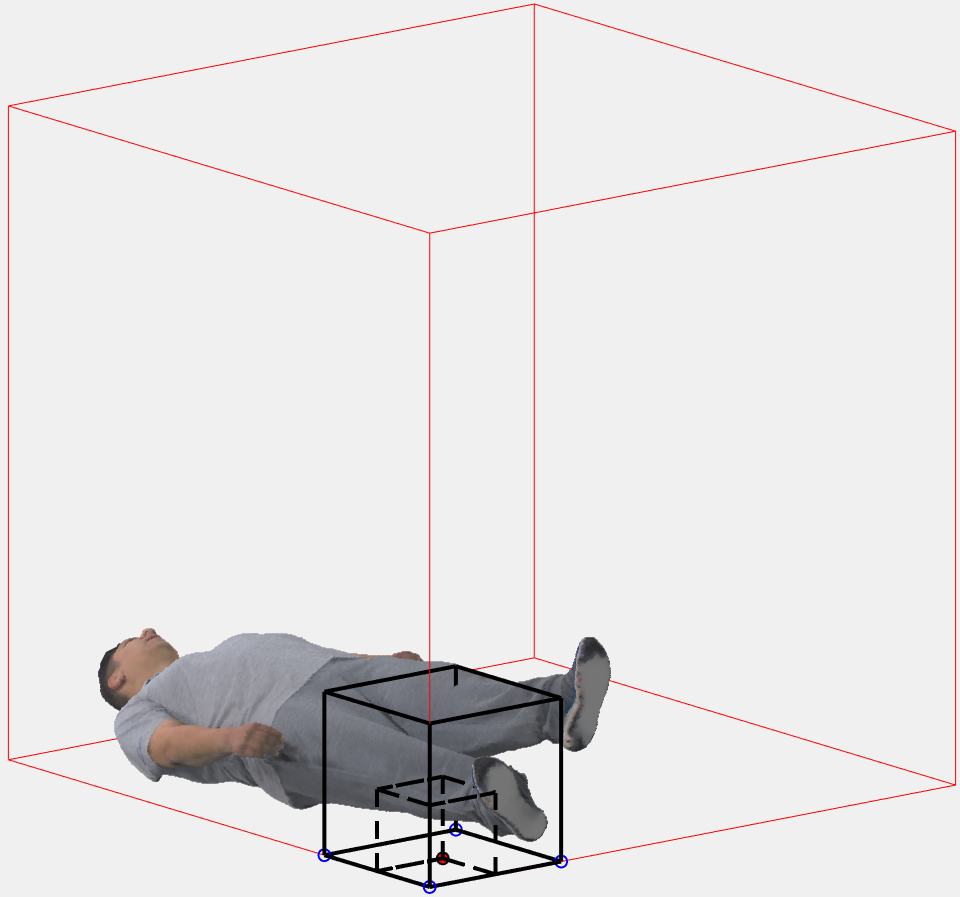}
\includegraphics[width=0.6\linewidth]{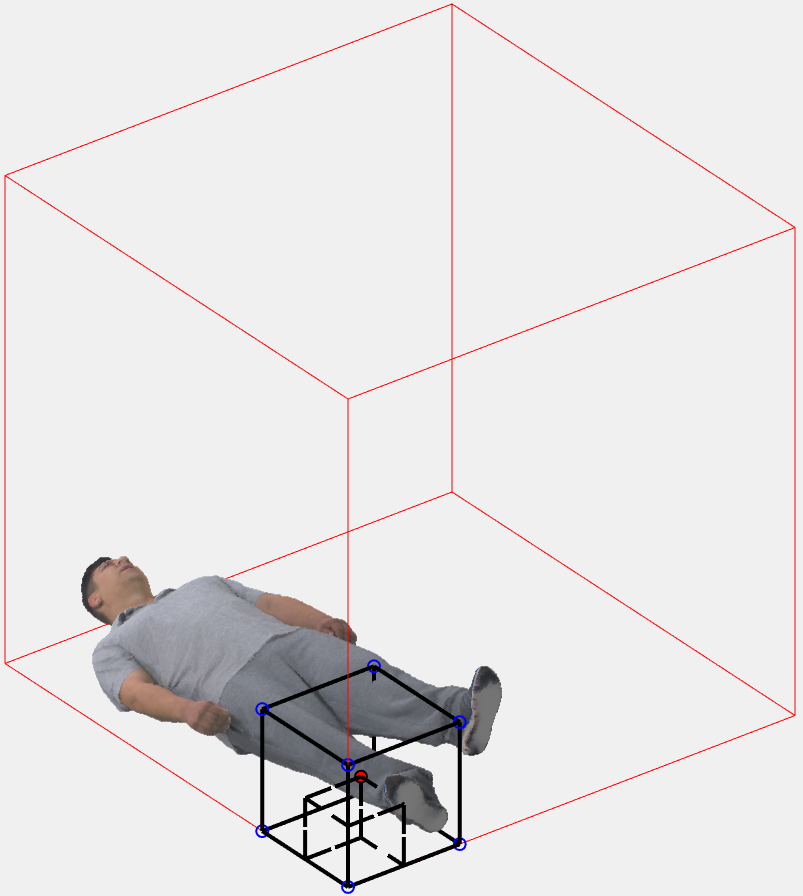}
\caption{Examples of parent corners at level $\ell$ used to obtain predictions for a child corner at level $\ell$ + 1.  The child corners are colored in red, the corresponding parent corners are colored in blue, and the root octree block (red) is shown for reference. (Top) The child corner is on a parent block $edge$; (Middle) The child corner is on a parent block $face$; (Bottom) The child corner is in the $center$ of the parent block.}
\label{fig:SDFPredictions}
\end{figure}

Fig.~\ref{fig:SDFPredictions} shows examples of how we compute the SDF (control point) predictions for child corners at level $\ell$ + 1, depending on their location on the parent octree block at level $\ell$, i.e., whether the child corner is on a parent block $edge$, $face$, or in the parent block $center$.  The prediction is equal to the average of the SDF values of the parent corners circled in blue in Fig.~\ref{fig:SDFPredictions}, for the child corner colored in red in the same diagram.  The {\em prediction error}, or {\em residual}, or {\em wavelet coefficient}, for that child corner is then the difference between the child's SDF value and the prediction (average of the parent control points). 

\subsection{Octree Coding of B\'ezier Volumes}

To compress an implicit surface $\{\bm{x}:f(\bm{x})=c\}$, we compress $f$ to a function $\hat f$ and represent the surface as $\{\bm{x}:\hat f(\bm{x})=c\}$, where $c=0$.

The straightforward way to compress $f$ would be to project it to a function $f_\ell$ with sufficiently high level of detail $\ell=d$, represent $f_d$ as the sum of functions $f_0+g_0+g_1+\cdots +g_{d-1}$, and then quantize and entropy code the coefficients $F_{0,\bm{n}}$ (for all $\bm{n}\in\mathcal{C}_0$), $G_{0,\bm{n}}$ (for all $\bm{n}\in\mathcal{C}_1\setminus2\mathcal{C}_0$), $G_{1,\bm{n}}$ (for all $\bm{n}\in\mathcal{C}_2\setminus2\mathcal{C}_1$), and so forth up through $G_{d-1,\bm{n}}$ (for all $\bm{n}\in\mathcal{C}_d\setminus2\mathcal{C}_{d-1}$).

However, because the subspaces $\mathcal{F}_0,\mathcal{G}_0,\mathcal{G}_1,\ldots,\mathcal{G}_{\ell-1}$ are not orthogonal to each other (in the least squares sense), in this straightforward approach, the norm of the quantization error would be larger in the reconstructed domain than in the coefficient domain, due to error propagation.

To mitigate the problem of error propagation, $f$ is compressed sequentially with quantization in the loop as follows.  Initially, $f$ is projected onto the subspace $\mathcal{F}_0$ by sampling $f$ on the corners of the unit cube, $F_{0,\bm{n}}=f_0(\bm{n})=f(\bm{n})$ (for all $\bm{n}\in\mathcal{C}_0$).  These low pass coefficients are uniformly scalar quantized with stepsize $\Delta$, entropy coded, and transmitted.  Let $\hat F_{0,\bm{n}}=\hat f_0(\bm{n})=\hat f(\bm{n})$ (for all $\bm{n}\in\mathcal{C}_0$) denote their reconstruction.  Then, for $\ell=0,1,\ldots,d-1$ in sequence, the prediction $\hat{f}_\ell(2^{-(\ell+1)}\bm{n})$ is formed by tri-linear interpolation of $\hat f_\ell$ (for all $\bm{n}\in\mathcal{C}_{\ell+1}$); $f$ is projected onto the subspace $\mathcal{F}_{\ell+1}$ as $F_{\ell+1,\bm{n}}=f_{\ell+1}(2^{-(\ell+1)}\bm{n})=f(2^{-(\ell+1)}\bm{n})$ (for all $\bm{n}\in\mathcal{C}_{\ell+1}$); the wavelet coefficients $G_{\ell,\bm{n}}=F_{\ell+1,\bm{n}}-\hat{f}_\ell(2^{-(\ell+1)}\bm{n})$ are uniformly scalar quantized with stepsize $\Delta$, entropy coded, transmitted, and recovered as $\hat G_{\ell,\bm{n}}$ (for all $\bm{n}\in\mathcal{C}_{\ell+1}\setminus2\mathcal{C}_\ell$); and the quantized low pass coefficients at level $\ell+1$ are recovered as $\hat F_{\ell+1,\bm{n}}=\hat G_{\ell,\bm{n}}+\hat{f}_\ell(2^{-(\ell+1)}\bm{n})$ (for all $\bm{n}\in\mathcal{C}_{\ell+1}$).  This guarantees that \begin{eqnarray}
\lefteqn{|f_{\ell+1}(2^{-(\ell+1)}\bm{n})-\hat f_{\ell+1}(2^{-(\ell+1)}\bm{n})|} \label{eqn:Delta_guarantee} \\
& = & |F_{\ell+1,\bm{n}}-\hat F_{\ell+1,\bm{n}}|=|G_{\ell+1,\bm{n}}-\hat G_{\ell+1,\bm{n}}|\leq\Delta/2 \nonumber
\end{eqnarray}
for all $\ell$ and $\bm{n}\in\mathbb{Z}^3$.

If $f$ is the signed distance function, then (\ref{eqn:Delta_guarantee}) guarantees that the location of the reconstructed surface $\{\bm{x}:\hat f(\bm{x})=0\}$ is within Hausdorff distance $\Delta/2$ of the original surface $\{\bm{x}:f(\bm{x})=0\}$.

\subsection{Pruning the Octree Before Encoding}

It is usually not necessary to transmit the entire octree and all of its associated wavelet coefficients to the decoder, particularly in the case of {\em lossy} shape compression.  One reason for this is that many of the quantized wavelet coefficients will be zero, or near zero, close to the leaves (voxel level) of the octree, or wherever the surface is locally flat (or nearly flat).  This means that the octree blocks with such small wavelet coefficients will not contribute much to the overall quality of shape reconstruction of the 3D object.  We therefore experimented with several different methods for octree pruning, all of which fit into the following two categories:

\begin{itemize}
\item {\bf Fixed-level pruning:} All the octree blocks and their associated wavelet coefficients beyond one chosen octree level are pruned off.  In this case, all the octree {\em leaves} (blocks with no descendants) end up being at the same octree level after pruning.

\item {\bf Variable-level pruning:} Octree blocks and their associated wavelet coefficients are pruned according to some rate-distortion algorithm or other criteria applied on different branches of the octree, so that after pruning the octree leaves can be at variable octree levels.
\end{itemize}
The octree {\em leaves} that remain after pruning in any of the methods described above are termed the {\em B\'ezier Volumes} (BVs) in our work.  

For the case of variable-level pruning, we also tried a number of different solutions, mainly:
\begin{itemize}
\item {\em Pruning based on zero wavelet coefficients:} When the wavelet coefficients are zero on the corners of all of a block's descendant blocks, then the octree may be pruned at that ancestral block, leaving the ancestral block as a leaf, or {\em B\'ezier Volume} (BV) of order $p=2$, so-called because the value of $\hat f$ within the block can be completely determined by tri-linear interpolation of the values of the low-pass coefficients, or {\em control points}, at the corners of the BV block.  Thus the surface $\{\bm{x}:\hat f(\bm{x})=0\}$ within the block is likewise determined by tri-linear interpolation of the control points on the corners of the block.  Furthermore, because neighboring blocks (even if they are at different levels) share the same values on any shared corners, they share the same values on any shared face; hence they share the same level set on any shared face; hence the reconstructed surface between neighboring blocks is continuous regardless of the quantization stepsize $\Delta$.

\item {\em Pruning based on zero and small non-zero wavelet coefficients:} This is the same method as described above, except that instead of only considering quantized wavelet coefficients with values of zero, we also consider quantized wavelet coefficients with other "small" (within some chosen threshold) non-zero values.  This effectively corresponds to applying a deadzone to the uniform scalar quantizer that is used to quantize the wavelet coefficients.  

\item {\em Pruning based on estimated geometry reconstruction error:} The descendants of octree blocks in which the estimated geometry reconstruction error is "small enough" (i.e., within some chosen error threshold) are pruned off, along with their associated wavelet coefficients.  

\item {\em Pruning based on optimal rate-distortion theory:} Instead of only considering the estimated geometry reconstruction error as described in the method above, in this case we also consider the estimated changes in bitrate if an octree block and its descendants are pruned versus not pruned.  This method is based on the optimal rate-distortion pruning theory described in \cite{ChouLG:89}.
\end{itemize}
The experimental results and discussions on the successes and failures of the pruning methods described above are provided in section VI.B.

\subsection{Reconstructing a point cloud from the implicit surface}

The ability to reconstruct a volumetric function $\hat f$ whose level set represents the surface implicitly is not usually the end result.  In systems of practical interest, the surface must be made explicit.  Typically, this means rasterizing or voxelizing the implicit surface into a finite set of points.  This can be considered {\em rendering} or {\em reconstructing} an explicit point cloud from the implicit surface.

In this subsection, we outline two methods in which the implicit surface within a B\'ezier Volume can be made explicit.

In the first method, known as {\em recursive subdivision}, the B\'ezier Volume is recursively subdivided to a particular level.  We say that a block at level $\ell$ has a {\em $c$-crossing} if $\min\{\hat F_{\ell,\bm{n}}\}\leq c\leq\max\{\hat F_{\ell,\bm{n}}\}$, where the min and max are over all eight corners $\bm{n}$ of the block.  That is, some reconstructed control points lie on one side of $c$, and some lie on the other.  If a block has a $c$-crossing, then it contains a part of the implicit surface $\{\bm{x}:\hat f(\bm{x})=c\}$, and hence is {\em occupied}.  An occupied block at level $\ell=d$ is declared to be an occupied voxel, while an occupied block at level $\ell<d$ is subdivided into eight subblocks at level $\ell+1$, and the process is repeated on each occupied subblock after computing the control points of the subblock using tri-linear interpolation of the control points of the parent block.  This procedure is guaranteed to find all occupied voxels.

In the second method, known as {\em ray-casting}, the B\'ezier Volume is raster-scanned down the dominant axis of the surface within the BV.  The dominant axis of the surface is the one closest to an estimated surface normal.  We estimate the surface normal by taking the gradient of the volumetric function $\hat f$ within the BV.  Assume the BV is normalized to the unit cube $[0,1]^3$, and that the control points at its eight corners are $\hat F(i,j,k)$, $i,j,k\in\{0,1\}$.  Then at any point $(x,y,z)$ within the BV, the value of $\hat f(x,y,z)$ is tri-linear interpolated as
\begin{eqnarray}
\hat f(x,y,z) & = & \sum_{i=0}^1 \sum_{j=0}^1 \sum_{k=0}^1
F(i,j,k) \\
& & x^i(1-x)^{1-i} y^j(1-y)^{1-j} z^k(1-z)^{1-k}. \nonumber
\end{eqnarray}
Thus the gradient within the BV has constant components
\begin{eqnarray}
\frac{\partial\hat f}{\partial x}
& = & \sum_{j=0}^1 \sum_{k=0}^1 [F(1,j,k)-F(0,j,k)] \\
\frac{\partial\hat f}{\partial y}
& = & \sum_{i=0}^1 \sum_{k=0}^1 [F(i,1,k)-F(i,0,k)] \\
\frac{\partial\hat f}{\partial z}
& = & \sum_{i=0}^1 \sum_{j=0}^1 [F(i,j,1)-F(i,j,0)].
\end{eqnarray}
The component whose absolute value is the largest determines the dominant axis.  Say the dominant axis is $z$.  Then the $(x,y)$ plane is rasterized into a finite set of points $(x_m,y_m)\in[0,1]^2$, and for each $(x_i,y_i)$ the intersection with the implicit surface by the ray through $(x_i,y_i)$ along $z$ is determined as
\begin{align}
z_m & = & \left[\sum_{i=0}^1 \sum_{j=0}^1 F(i,j,0) x_m^i(1-x_m)^{1-i} y_m^j(1-y_m)^{1-j}\right]  \nonumber \\
& / & \left[\sum_{i=0}^1 \sum_{j=0}^1 F(i,j,0) x_m^i(1-x_m)^{1-i} y_m^j(1-y_m)^{1-j}\right. \nonumber \\
& - & \left.\sum_{i=0}^1 \sum_{j=0}^1 F(i,j,1) x_m^i(1-x_m)^{1-i} y_m^j(1-y_m)^{1-j}\right].
\end{align}
The collection of points $\{(x_m,y_m,z_m)\}$ is thus a rasterization of the implicit surface.

\section{Experimental Results}
\label{sec:results}

\subsection{Attribute Coding Results}

The framework presented in this paper allows us to extend RAHT from order $p=1$ to more general B\'ezier Volumes $p\geq 1$.  In this section, we show the improvement of $p=2$ over $p=1$ for attribute representation and compression.  We assume the geometry is known, and examine only color attributes.  Evaluations are performed on the datasets listed in Table~\ref{tab}.  These datasets are the ones used in the MPEG Point Cloud Coding standardization activity to compare attribute coding methods, assuming lossless geometry.  The original datasets all have 10-bit resolution, with the exception of {\em shiva}, which has 12-bit resolution.  ($d$-bit resolution means that the point locations are represented as $d$-bit integers.)  For our experiments, all original datasets are first voxelized to either 10-bit or 7-bit resolution.

\begin{table}
\centering
\begin{tabular}{|r|r|r|r|r|} \hline
Dataset name & \parbox[b]{0.35in}{\raggedleft Original resolution} & \parbox[b]{0.35in}{\raggedleft Original point count} & \parbox[b]{0.35in}{\raggedleft 10-bit point count} & \parbox[b]{0.35in}{\raggedleft 7-bit point count} \\ \hline 
{\em loot\_vox10\_1200} & 10 bit & 805285 & 805285 & 14711 \\
{\em queen\_0200} & 10 bit & 1000993 & 1000993 & 14683 \\
{\em soldier\_vox10\_0690} & 10 bit & 1089091 & 1089091 & 19993 \\
{\em shiva\_00035\_vox12} & 12 bit & 1009132 & 900662 & 30045 \\
{\em longdress\_vox10\_1300} & 10 bit & 857966 & 857966 & 15688 \\ \hline
\end{tabular}
\linebreak
\caption{Datasets used for Attribute representation and compression}
\label{tab}
\end{table}

\begin{figure*}[!t]
\centering
\vspace*{-0.25in}
\includegraphics[width=0.18\linewidth]{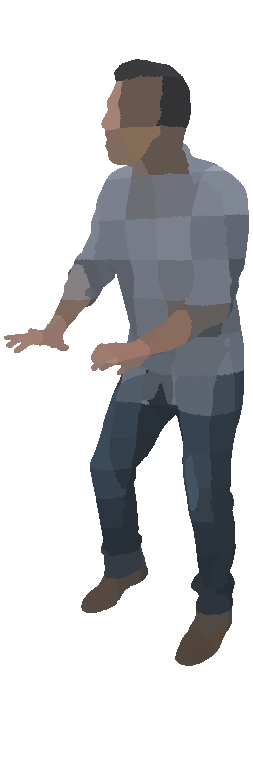}
\includegraphics[width=0.18\linewidth]{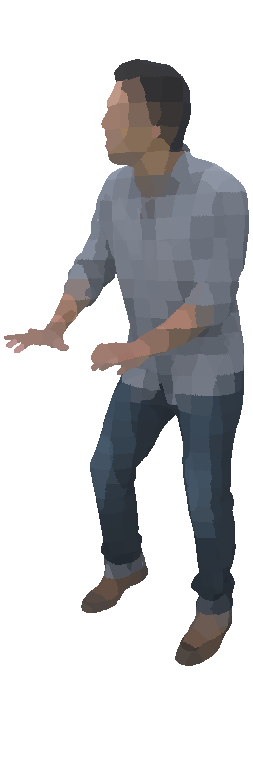}
\includegraphics[width=0.18\linewidth]{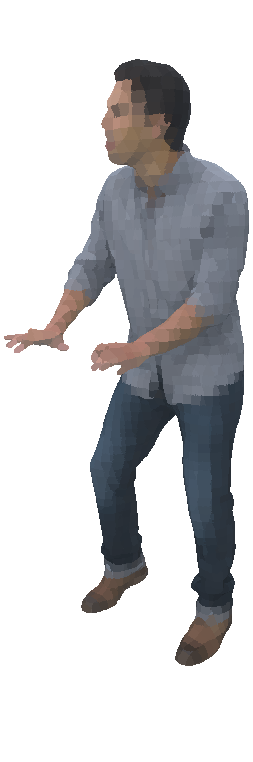}
\includegraphics[width=0.18\linewidth]{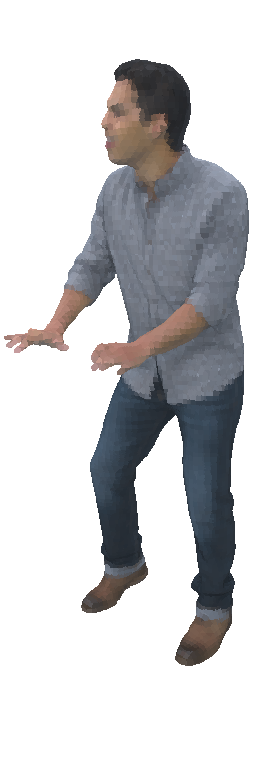}
\includegraphics[width=0.18\linewidth]{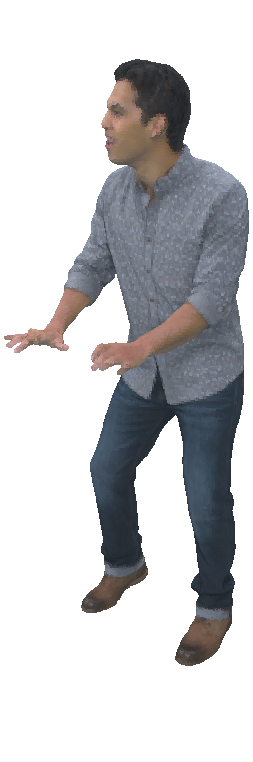}

\vspace*{-0.75in}
\includegraphics[width=0.18\linewidth]{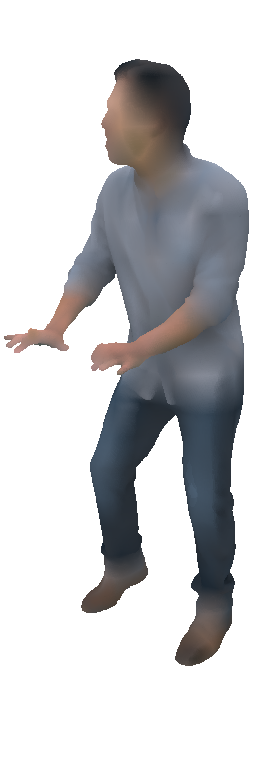}
\includegraphics[width=0.18\linewidth]{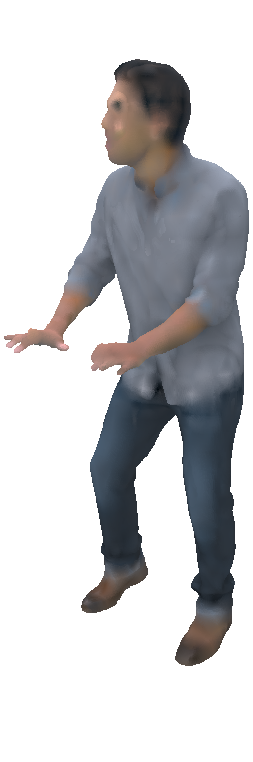}
\includegraphics[width=0.18\linewidth]{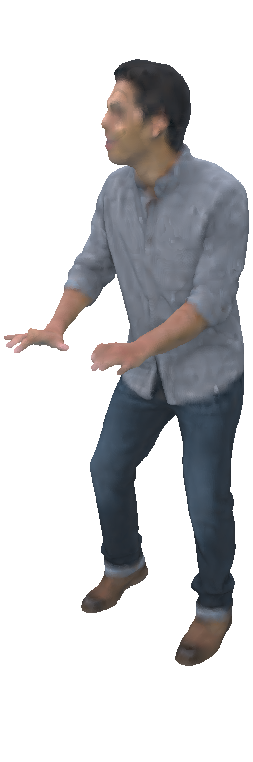}
\includegraphics[width=0.18\linewidth]{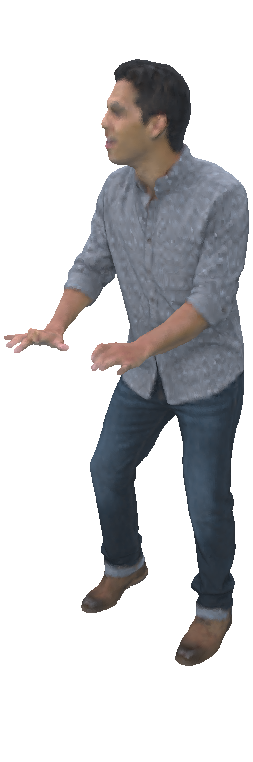}
\includegraphics[width=0.18\linewidth]{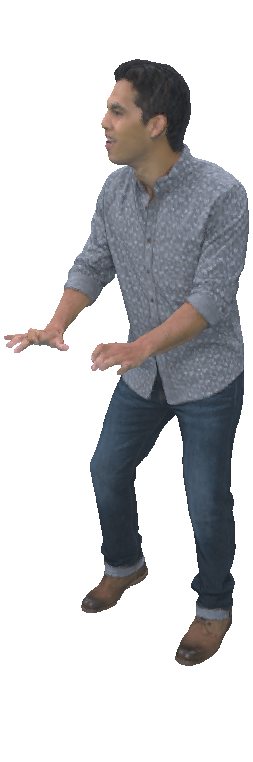}

\vspace*{-0.5in}
\caption{Smoothing the {\em loot} dataset for $p=1$ (RAHT, top) and $p=2$ (BV, bottom), at levels 4 (left), 5, 6, 7, and 8 (right).}
\label{fig:RAHTvsBVsmoothing}
\end{figure*}

Fig.~\ref{fig:RAHTvsBVsmoothing} shows the qualitative results of smoothing the 10-bit {\em loot} dataset by setting the wavelet coefficients $G_\ell$ to zero for all levels $\ell\geq L$, where $L=4$ (left) through $L=8$ (right), for $p=1$ (RAHT, top) and $p=2$ (BV, bottom).  The corresponding blocks have blockwidths $2^{10-L}$, i.e., the blocks are $64\times64\times64$ voxels (left) through $4\times4\times4$ voxels (right). For RAHT, blocking artifacts are clearly visible, while for BV, blocking artifacts are much less visible because the colors are guaranteed to be continuous between blocks, at every $L$.  The improvement from $p=1$ to $p=2$ is striking.

\begin{figure}[!t]
\centering
\includegraphics[width=1.0\linewidth]{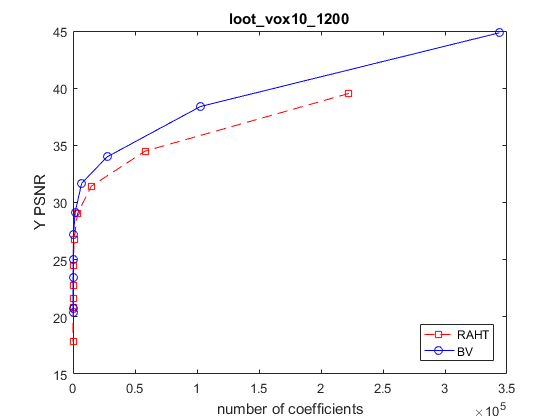}
\includegraphics[width=0.45\linewidth]{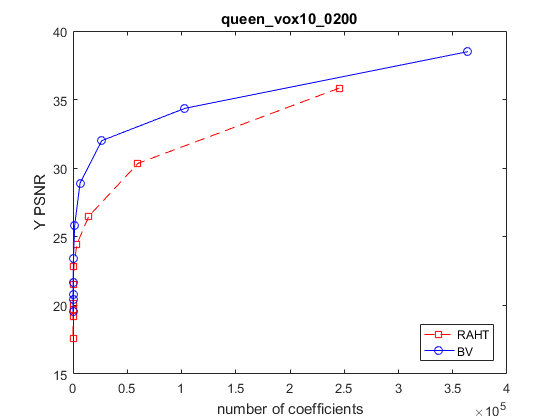}
\includegraphics[width=0.45\linewidth]{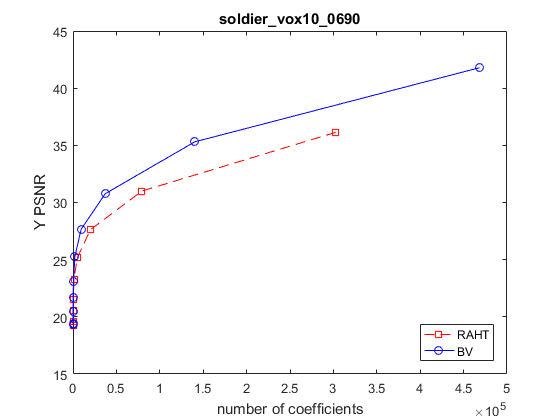}
\includegraphics[width=0.45\linewidth]{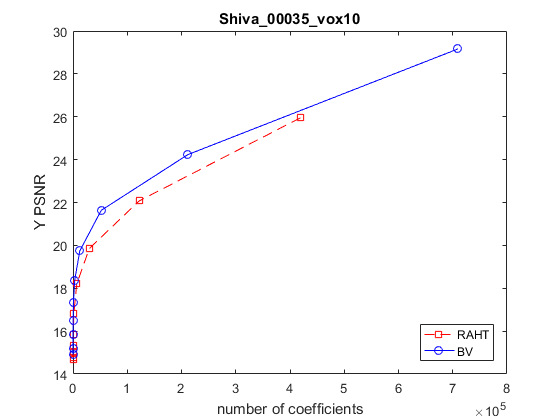}
\includegraphics[width=0.45\linewidth]{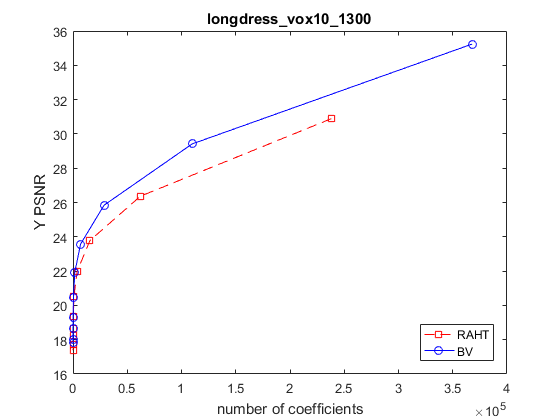}
\caption{Energy compaction for $p=2$ (BV) vs.\ $p=1$ (RAHT).}
\label{fig:RAHTvsBVenergyCompaction}
\end{figure}

Fig.~\ref{fig:RAHTvsBVenergyCompaction} shows the smoothing results quantitatively, as $L$ is swept from 0 to 9, on all five 10-bit datasets.  Each plot is a graph of Y (luma) PSNR as a function of the number of non-zero wavelet coefficients in levels $\ell<L$, for $p=1$ (RAHT, red dashed line with squares) and $p=2$ (BV, blue solid line with circles).  The gap between the lines indicates the improvement of BV over RAHT in energy compaction gain, or transform coding gain, about 3-4 dB.

\begin{figure}[!t]
\centering
\includegraphics[width=1.0\linewidth]{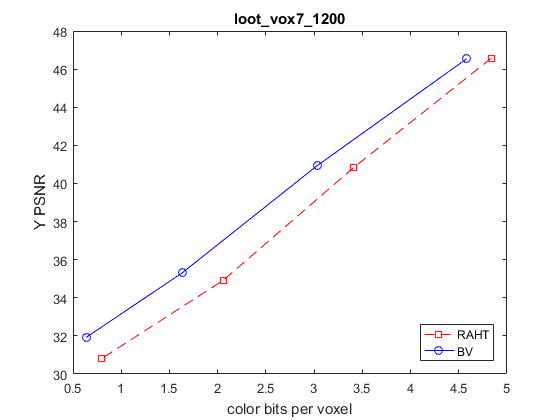}
\includegraphics[width=0.45\linewidth]{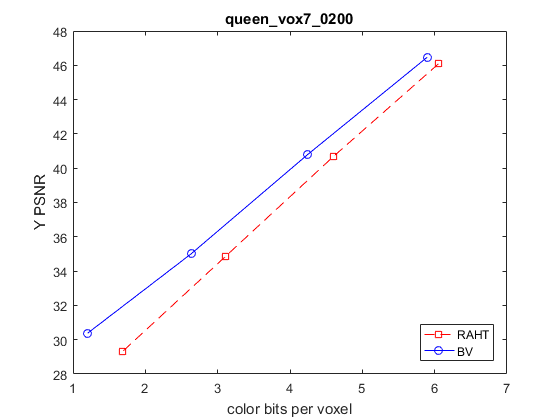}
\includegraphics[width=0.45\linewidth]{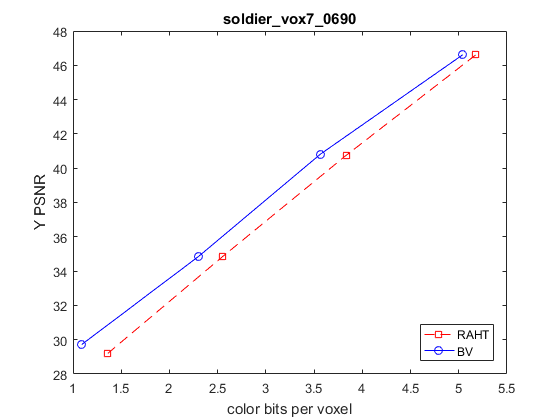}
\includegraphics[width=0.45\linewidth]{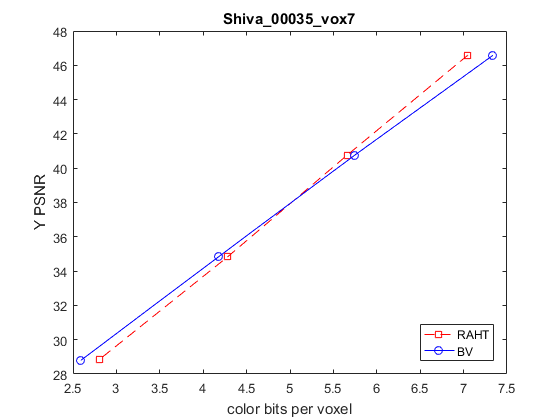}
\includegraphics[width=0.45\linewidth]{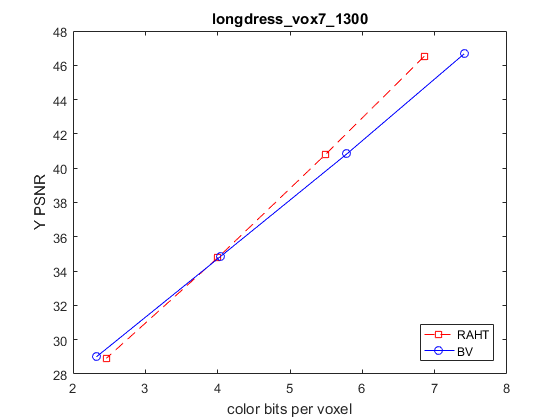}
\caption{Distortion-rate performance for $p=2$ (BV) vs.\ $p=1$ (RAHT).}
\label{fig:RAHTvsBVrateDistortion}
\end{figure}

Fig.~\ref{fig:RAHTvsBVrateDistortion} shows how the improvement in energy compaction gain translates into actual coding gain, on all five 7-bit datasets.  Each plot is a graph of Y PSNR as a function of the total color (Y+U+V) bitrate.  Here, for entropy coding, we use adaptive Run-Length Golomb-Rice (RLGR) coding (\cite{malvar_rlgr}) separately on each color component (Y, U, V) on each level.  Though this is the preferred method of entropy coding for RAHT (\cite{SandriRC:18}), as of yet there has been no study of entropy coding for BV with $p\geq2$, so potentially the results for BV may improve (e.g., by estimating the variance of each coefficient and using it as context for the entropy coder).  Regardless, the plots show that the improvement of BV over RAHT in distortion for a given bit rate can be close to 2 dB, especially at lower bit rates.  Datasets with smoother color textures ({\em loot} and {\em queen}) show larger gains than datasets with medium color textures ({\em soldier}).  Datasets with high-frequency color texture ({\em shiva} and {\em longdress}) demonstrate little coding gain of BV over RAHT, despite still having a significant energy compaction gain.  The reason probably has to do with a mismatched entropy code, but requires more investigation.

\begin{table}
\centering
\begin{tabular}{|r|r|r|r|r|} \hline
Dataset name & \parbox[b]{0.8in}{\raggedleft Original resolution} & \parbox[b]{0.8in}{\raggedleft Original point count}\\ \hline 
{\em boxer\_8i\_vox10} & 10 bit & 993745\\
{\em longdress\_1300\_8i\_vox10} & 10 bit & 911432\\
{\em loot\_1200\_8i\_vox10} & 10 bit & 868658\\
{\em soldier\_0690\_8i\_vox10} & 10 bit & 1191745\\ \hline
\end{tabular}
\linebreak
\label{tab2}
\caption{Datasets used for Geometry representation and compression}
\end{table}

\subsection{Geometry Coding Results}

The results presented in this section relate to the datasets in Table II.  All of the point clouds in Table II have 10-bit geometry resolution, meaning that the point ($x$, $y$, $z$) locations are integers in the range [0, 1023].  The $longdress$, $loot$, and $soldie$r datasets are similar to the corresponding datasets used by the MPEG Point Cloud Coding (PCC) group, except that the normals have been recomputed to be more accurate and the point count is slightly different.  The 10-bit version of the $boxer$ dataset is not currently used by the MPEG PCC group.

\begin{figure}[!t]
\centering
\includegraphics[width=0.8\linewidth]{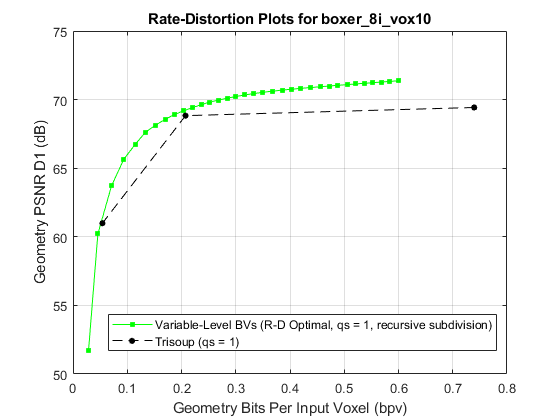}
\includegraphics[width=0.8\linewidth]{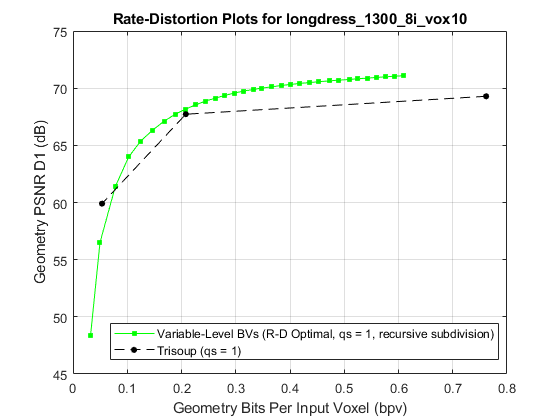}
\includegraphics[width=0.8\linewidth]{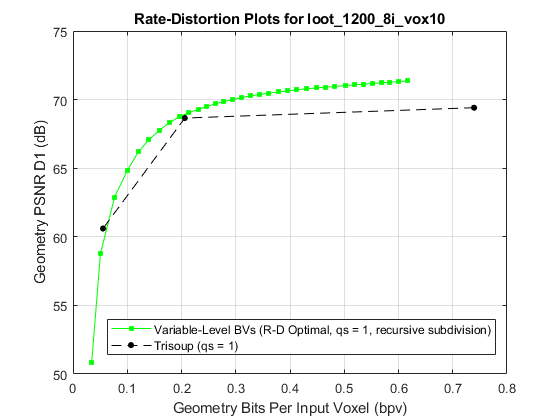}
\includegraphics[width=0.8\linewidth]{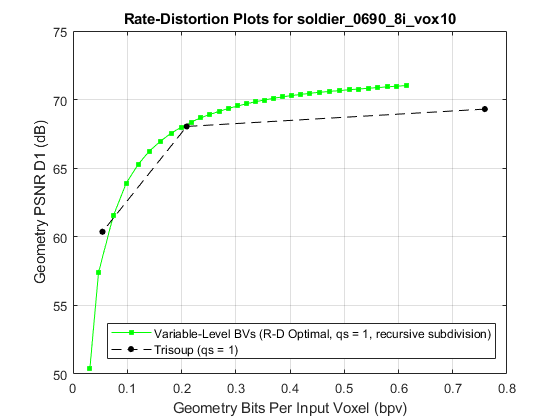}
\caption{Rate-distortion performance for BV vs Trisoup, when BV uses rate-distortion optimal pruning and recursive subdivision for voxel reconstruction.}
\label{fig:BestBVvsTrisoupRD}
\end{figure}

Fig.~\ref{fig:BestBVvsTrisoupRD} shows the rate-distortion performance of the BV geometry compression method versus the Trisoup (or S-PCC) method that is the Test Model for geometry compression of static (single-frame) point clouds in the emerging MPEG G-PCC standard \cite{Schwarz:18}.  For the BV method in Fig.~\ref{fig:BestBVvsTrisoupRD}, the pruning method used is the {\em rate-distortion optimal} pruning (see section V.D.), and the {\em recursive subdivision} method (see section V.E.) is used to reconstruct the voxels from the implicit surface. The {\em rate} in the plots in Fig.~\ref{fig:BestBVvsTrisoupRD} represents the geometry bitrate measured as bits per input voxel of the corresponding input point cloud, where the number of input voxels is as shown in Table II.  Included in the geometry bitrate for BV are the following: 
\begin{itemize}
	\item Occupancy codes of the pruned octree, using 8 bits per internal (non-leaf) octree node and compressed using 7-zip,
	\item A ``post-pruning array'' that indicates which of the octree blocks that remain after pruning are leaves (represented by a ``1'' bit) and which are internal (represented by a ``0'' bit), packed into bytes and compressed using 7-zip,
    \item Control points (SDF values) at a chosen coarse level (``start level'') of the octree (level 2 was chosen for the results presented in this paper), uniformly scalar-quantized and compressed using 7-zip,
    \item Wavelet coefficients at each octree level, from the next level after "start level" up to and including the variable leaf levels (excluding the case where the leaves are at the voxel level, as these voxel positions are simply reconstructed from the occupancy codes and so do not need their SDFs to be reconstructed at the decoder), uniformly scalar-quantized and rANS entropy-encoded \cite{Duda:14} by octree level, then further compressed using 7-zip.
\end{itemize}
The individual 7-zip files obtained above are then all added to the same 7-zip archive and consolidated into one binary 7-zip file, whose size is used to obtain the BV geometry bitrates presented in this paper.

The {\em distortion} in Fig.~\ref{fig:BestBVvsTrisoupRD} is the point-to-point geometry PSNR obtained from the {\em pc\_error} measurement tool developed by the MPEG PCC group \cite{Schwarz:18b} (version 0.10 of pc\_error was used to obtain the geometry PSNR results in this paper).  The ``qs = 1'' for both the BV and Trisoup methods in Fig.~\ref{fig:BestBVvsTrisoupRD} indicates that a uniform scalar quantization stepsize of 1 was used for all the transmitted geometry-related data that was quantized.  

\begin{figure*}[!t]
\centering
\includegraphics[width=1.0\linewidth]{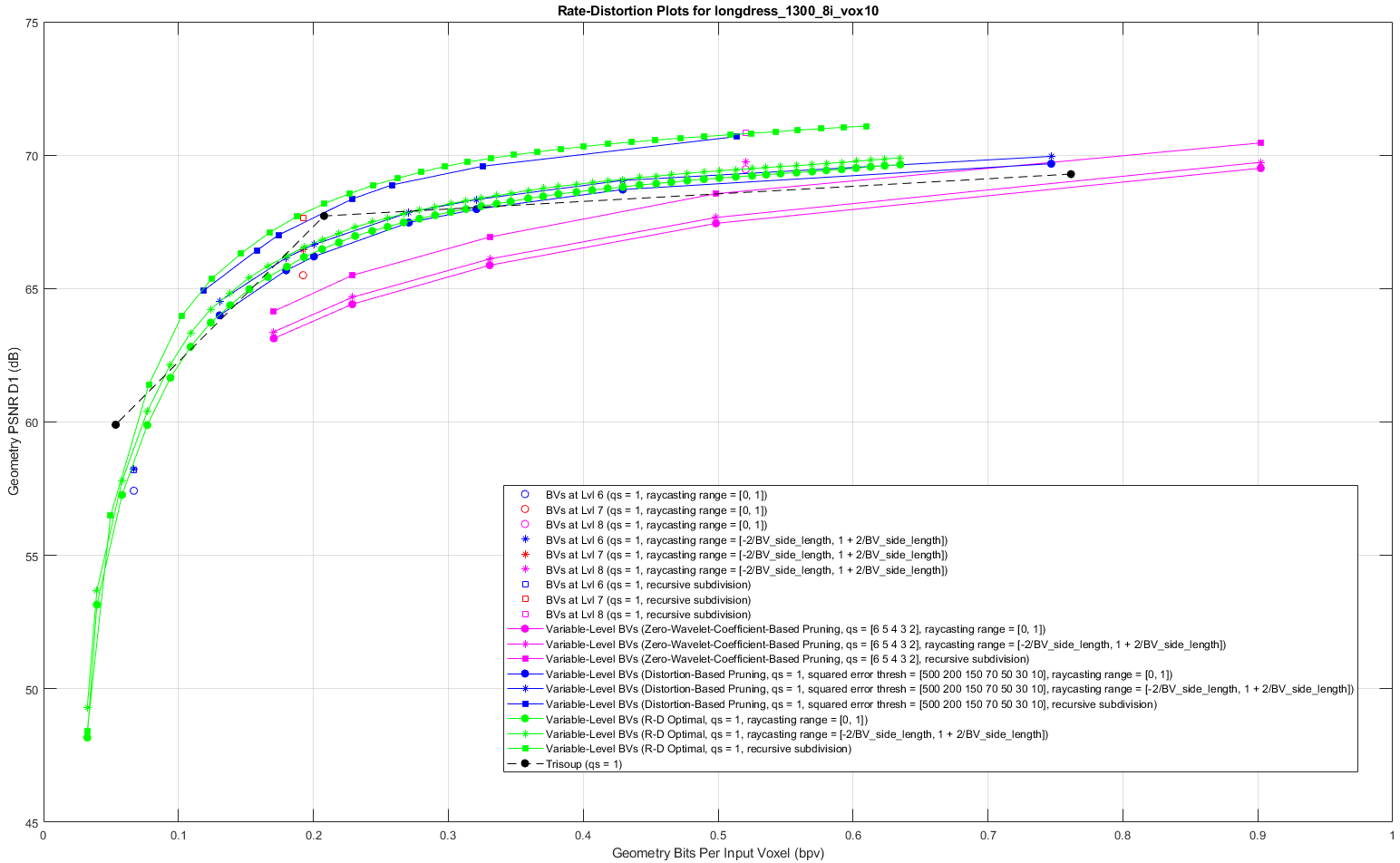}
\linebreak
\linebreak
\includegraphics[width=1.0\linewidth]{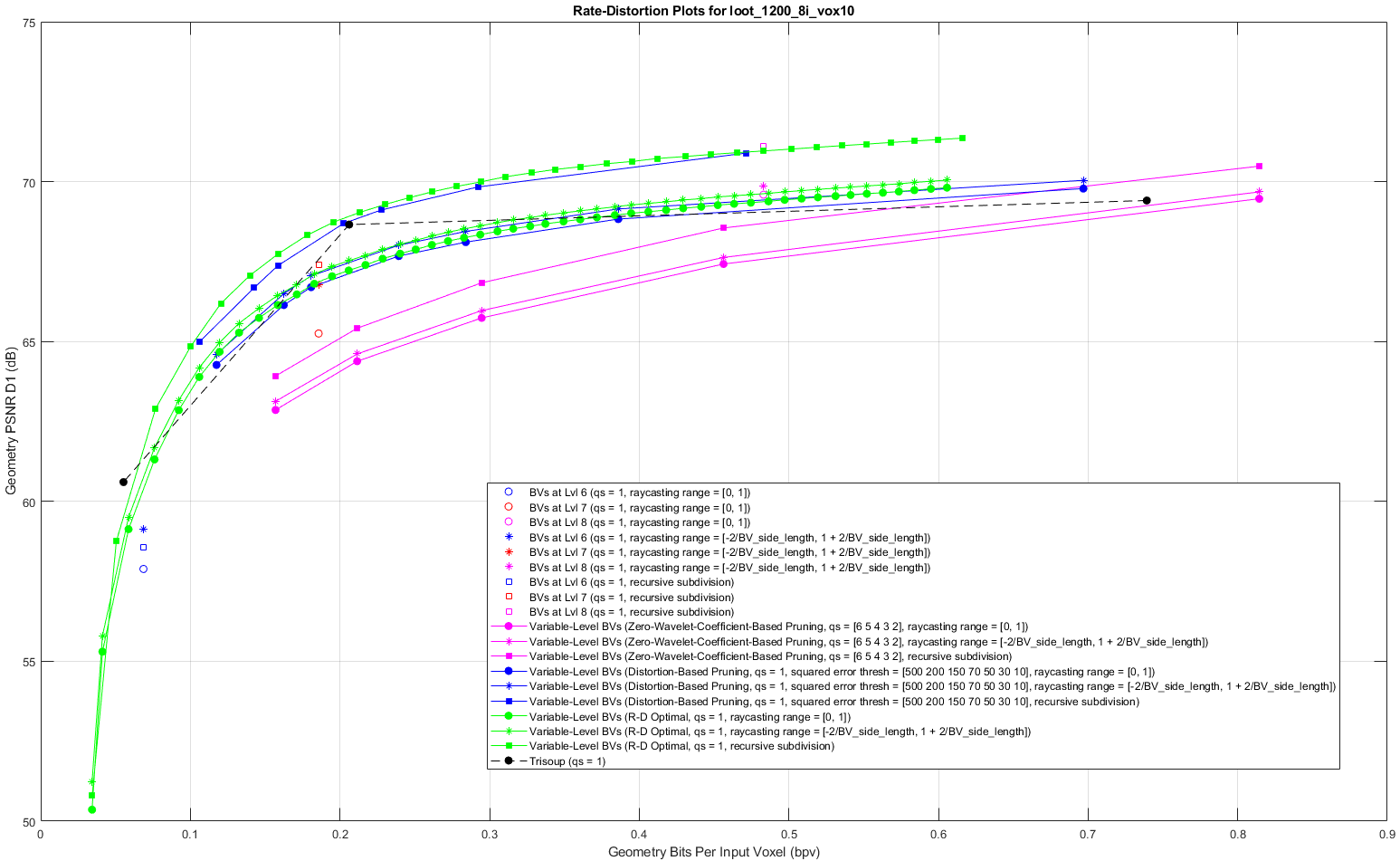}
\linebreak
\linebreak
\caption{Rate-distortion performance for BV vs Trisoup, when different pruning methods and voxel reconstruction methods are used for BV.}
\label{fig:AllBVvsTrisoupRD}
\end{figure*}

Fig.~\ref{fig:BestBVvsTrisoupRD} demonstrates that the BV with R-D-optimal pruning and recursive subdivision voxel reconstruction outperforms Trisoup at all, or almost all, bitrates, and achieves near-maximum geometry PSNR (around 70 dB) at less than half the bitrate that Trisoup requires to get close to that quality level (around 69 dB).  We have found that using this pruning method, combined with the recursive subdivision voxel reconstruction method, produces the best rate-distortion results for BV.  Fig.~\ref{fig:AllBVvsTrisoupRD} additionally compares the rate-distortion performance of the BV method when using the different pruning methods described in section V.D., combined with either the {\em recursive subdivision} or the {\em raycasting} method for voxel reconstruction (described in section V.E.).  The results in Fig.~\ref{fig:AllBVvsTrisoupRD} are generally representative of the R-D results for all 4 datasets in Table II.

Several observations may be made from the results in Fig.~\ref{fig:AllBVvsTrisoupRD}.  Firstly, it is clear that the {\em recursive subdivision} voxel reconstruction method usually leads to significantly better PSNR results than when {\em raycasting} is used with the same pruning method.  The wider raycasting range of [-2\slash BV\_side\_length, 1 + 2\slash BV\_side\_length] in Fig.~\ref{fig:AllBVvsTrisoupRD}, where {\em BV\_side\_length} represents the width of an octree leaf block (or BV), indicates that here the voxel reconstruction is allowed to go slightly outside the BV blocks (i.e., outside the usual raycasting range of [0, 1] (see section V.E.)).  We believe that the reason that the subdivision method outperforms raycasting, and the raycasting with a wider range outperforms the raycasting with the [0, 1] range, is that the former two methods produce a thicker voxelized surface.  This potentially allows for more accurate nearest neighbors to be found during the point-to-point PSNR computation in pc\_error, which matches input and output voxels based on a nearest-neighbor computation.  However, we have also found that increasing the raycasting range beyond about [-2.5\slash BV\_side\_length, 1 + 2.5\slash BV\_side\_length], while still improving the PSNR slightly over the lower raycasting ranges (but still not better than when using recursive subdivision), tends to produce a voxelized surface that is noticeably too thick in certain places, at least in the kinds of point clouds that we tested on.  That is, the extra voxels noticeably protrude out from the rest of the surface, which is visually unappealing.  This does not happen in the recursive subdivision case, which still achieves superior PSNR results. 

Fig.~\ref{fig:AllBVvsTrisoupRD} also shows that, as expected, the rate-distortion optimal pruning usually produces better results than the corresponding fixed-level pruning when using the same voxel reconstruction method.  However, it can sometimes happen that the fixed-level pruning has a slightly better PSNR, for example the fixed-level pruning at level 8 using raycasting is slightly better than the corresponding rate-distortion optimal pruning for $longdress$ and $loot$ in Fig.~\ref{fig:AllBVvsTrisoupRD}.  We believe that this can be attributed to the slight inaccuracies in the distortion and/or rate estimations during the rate-distortion optimal pruning.

\begin{figure*}[!t]
\centering
\vspace*{-6cm}
\includegraphics[width=0.18\linewidth]{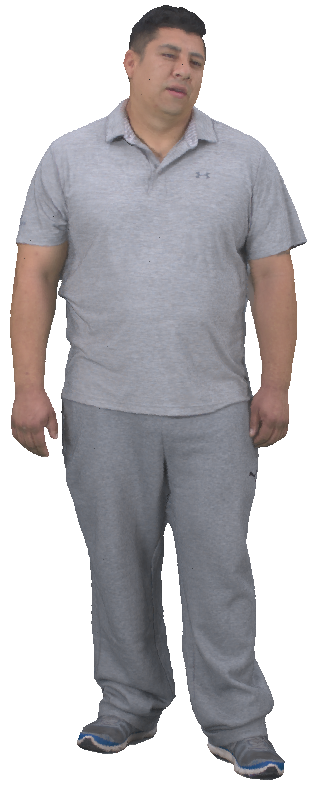}
\includegraphics[width=0.18\linewidth]{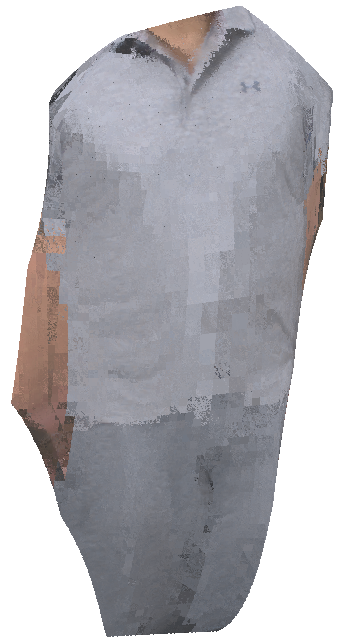}
\includegraphics[width=0.165\linewidth]{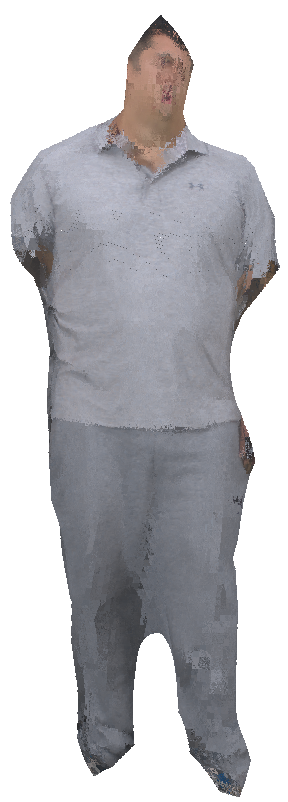}
\includegraphics[width=0.18\linewidth]{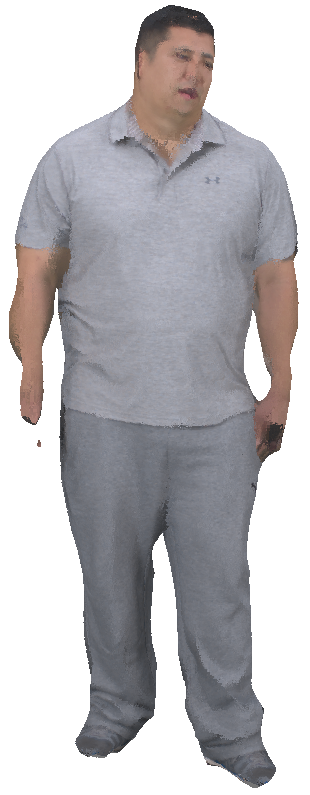}
\linebreak
\linebreak
\hspace*{3.5cm}
\includegraphics[width=0.175\linewidth]{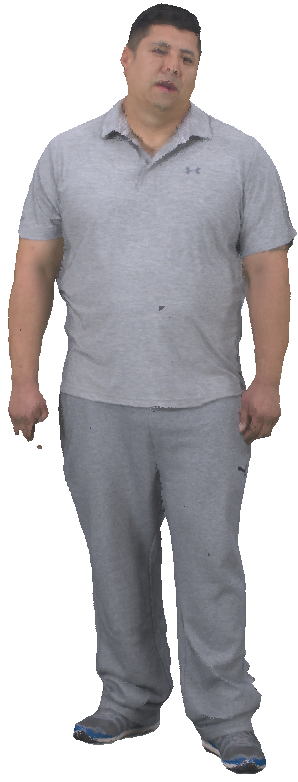}
\includegraphics[width=0.18\linewidth]{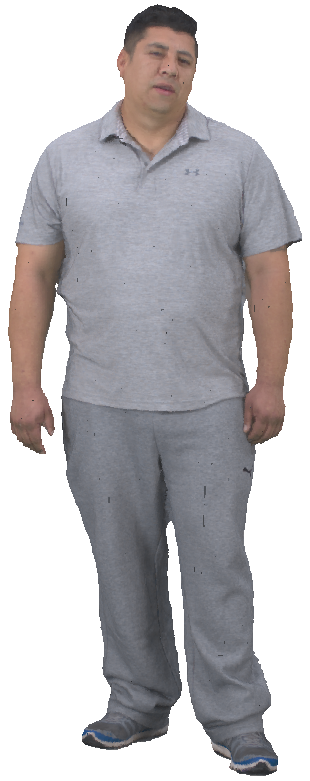}
\includegraphics[width=0.18\linewidth]{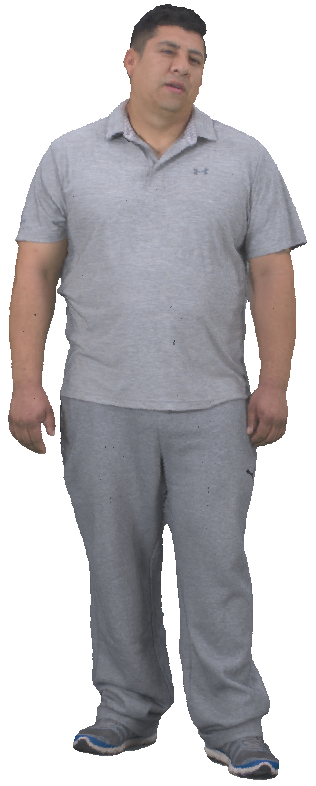}
\caption{Geometry (shape) reconstructions obtained with BV when using {\em fixed-level} pruning and {\em recursive subdivision} voxel reconstruction.  (Top leftmost) Original {\em boxer\_8i\_vox10} point cloud; (Top to bottom, left to right) Pruning levels: 3, 4, 5, 6, 7, 8.  Note that a {\em pruning level} here means that all the descendants of the octree blocks at that level are pruned off, so that the blocks at the pruning level become leaves (or BVs).  Also note that no color compression has been applied to the point clouds in the images above; the colors that are shown here have simply been mapped from the original point cloud by using a nearest-neighbor voxel matching.}
\label{fig:BoxerFixedLvlBVPruning}
\end{figure*}

Fig.~\ref{fig:BoxerFixedLvlBVPruning} shows examples of the geometry (shape) reconstructions obtained for {\em boxer\_8i\_vox10}, when {\em fixed-level} BV pruning is used with the {\em recursive subdivision} voxel reconstruction method.  We see that increasing the pruning level results in a progressively more refined shape reconstruction, and that pruning beyond octree level 8 (i.e., the BV blocks are at level 8) tends to produce an almost perfect geometry reconstruction, as confirmed by the PSNR results in Fig.~\ref{fig:AllBVvsTrisoupRD}.  Choosing a pruning level that is too small can result in BV blocks that are too big for the surface that fits inside them, so all of the control points on the corners of these BVs would have the same sign (+ in our implementation) since these corners are all outside the surface.  This means that, even though such a BV would be occupied, it would not contain a zero crossing, so it would not be subdivided during voxel reconstruction at the decoder; therefore, there would be no reconstructed surface (voxels) inside that block.  For example, this happens in the octree blocks at level 3 in Fig.~\ref{fig:PosSDFExample}, which contain the man's feet and his head.  This is why, in Fig.~\ref{fig:BoxerFixedLvlBVPruning}, for the pruning at level 3, the man has no head or feet (amongst other missing features). 

Probably the most surprising observation for us in Fig.~\ref{fig:AllBVvsTrisoupRD} is that the BV using variable-level octree pruning based on the locations of 0 wavelet coefficients has the worst performance, even being outperformed by the fixed-level pruning.  In fact, the fixed-level pruning performs much better than we initially expected, being very close to the rate-distortion optimal pruning.  The reason for the poor R-D performance of the variable-level BV based on 0 wavelet coefficients is probably due to the fact that, for the kinds of point clouds that we tested on, the 0 wavelet coefficients are usually only found very close to the voxel level.  This means that many of the original occupancy codes of the octree need to be transmitted, since the octree pruning does not go very deep into the tree, and these occupancy codes consume by far the largest proportion of the total geometry bitrate.  The R-D performance of this BV method can be somewhat improved by considering not only wavelet coefficients that have a value of 0 after quantization, but also other ``small'' wavelet coefficients (within some threshold), which is equivalent to applying a deadzone to the uniform scalar quantizer that is used to quantize the wavelet coefficients, as explained in section V.D..  However, we do not show these results in Fig.~\ref{fig:AllBVvsTrisoupRD}, because the performance gain is not significant enough to warrant ignoring the fact that with such a deadzone applied, the reconstructed SDF values are no longer guaranteed to be within +\slash- $\Delta$\slash2 of the original SDF values, where $\Delta$ is the quantization stepsize (see the explanation in section V.C.), and so unexpected artifacts can occur. 

The variable-level BV pruning that is based on considering estimated geometry reconstruction error only, without considering the estimated bitrate changes if an octree block is pruned or not, has a rate-distortion performance that is surprisingly close to the performance of the rate-distortion optimal pruning in Fig.~\ref{fig:AllBVvsTrisoupRD}.  Here the distortion estimation that is used within an octree block is the maximum of the two one-way squared errors between the reconstructed voxels in that block and their corresponding nearest neighbors in the same block in the original point cloud.  In Fig.~\ref{fig:AllBVvsTrisoupRD}, the different points on the corresponding R-D curves represent the different squared error values as defined in the legend (``squared error thresh''), where the largest error threshold (500) produces the lowest R-D point.  It is currently not clear why the performance of this pruning method is so close to the rate-distortion optimal pruning, but again, it might be related to inaccuracies in the rate/distortion estimation during the rate-distortion optimal pruning.  For example, the transmitted wavelet coefficients, which, in the R-D optimal pruning case consume most of the geometry bitrate at most of the rate points, are encoded at the end (after pruning) by using rANS entropy coding per octree level, and only encoding the wavelet coefficients for the {\em unique} octree block corners at each level, without repetitions for shared corners.  However, during rate-distortion-based pruning, the bitrates for the wavelet coefficients within a block are estimated using theoretical entropy estimates, not rANS, and currently there are no discounts in bits applied to account for the wavelet coefficients that are on shared octree corners, since we process one octree block at a time; so the overall bits estimate for the wavelet coefficients may be too high.

More generally, Fig.~\ref{fig:AllBVvsTrisoupRD} demonstrates that both the rate-distortion optimal BV pruning and the BV pruning based on estimated distortion only, combined with the recursive subdivision voxel reconstruction method, outperform the MPEG Trisoup method at most, if not all, bitrates.  Both BV and Trisoup are able to achieve a very good geometry reconstruction quality (around 69 dB) at the very small bitrate of approximately 0.2 bits per input voxel, or at an even smaller bitrate than that for BV. 

\section{Conclusion}
\label{sec:conclusion}

Volumetric media are just that: volumetric --- that is, defined as a signal on a set of points in 3D.  To date, the methods for processing volumetric media have been based on points or surfaces.  In this paper, we propose the first approach for processing both the geometry and attributes of volumetric media in a natively volumetric way: using volumetric functions.  Volumetric functions are functions defined everywhere in space, not just on the points of a point cloud.  Volumetric functions, especially if they are continuous, are well-suited to representing not only an original set of points in a point cloud and their attributes, but also nearby points and their attributes.  Just as regression fits continuous functions to a set of points and their values, and can therefore be used to interpolate the points and their values, volumetric functions can be used to interpolate the locations of points in a point cloud and their attributes.  Especially if they are continuous, this leads to, for example, infinitely zoomable geometry without holes, as well as spatially continuous attributes.  This is true regardless of the compression that may be applied to the volumetric functions, provided the compression preserves the continuity of the signal.

We choose to represent volumetric functions in a B-spline basis.  Though other choices could be made, the B-spline basis maps naturally to a multi-resolution, wavelet framework.  Detail is thus easily preserved where necessary, and smooth regions are processed efficiently.  Our multi-resolution approach fits perfectly with the octree approach for geometry processing.  In our approach, the octree can be considered a {\em non-zero tree} (which is in some sense the complement of the zero-tree used in image processing \cite{Shapiro:93}) in that it compactly codes the locations of non-zero (rather than zero) wavelet coefficients of the volumetric functions.

Beyond the idea of the volumetric functions for point clouds and the mechanics of coding them, our paper introduces two related, but distinct ideas.  The first is that geometry can be compressed in its implicit representation as a level set of a volumetric function.  The second is that one of the most successful and practical transforms previously used for point cloud attribute compression, namely RAHT, can be regarded as a continuous B-spline wavelet transform of order $p=1$, and thereby can be generalized to higher orders, which we relate to B\'ezier volumes, for higher performance.

Experimental results confirm, for both geometry and attribute compression of point clouds, that the volumetric approach is superior to previous approaches.


%



\section*{Acknowledgment}

The authors would like to thank their colleagues at 8i, Charles Loop, Robert Higgs, and Gianluca Cernigliaro, for inspiring discussions leading to the formulation of this work.

\ifCLASSOPTIONcaptionsoff
  \newpage
\fi



\bibliographystyle{IEEEtran}
\bibliography{bibtex/bib/IEEEabrv,bibtex/bib/IEEEexample}
%




%




\end{document}